\documentclass[
 reprint,
 amsmath,amssymb,
 aps,
]{revtex4-2}

\usepackage{comment}
\usepackage{lipsum}
\usepackage{appendix}
\usepackage{gensymb}
\usepackage{MnSymbol}
\usepackage{xcolor} 
\usepackage[abs]{overpic} 
\usepackage{graphicx}
\usepackage{subfig} 
\usepackage[justification=Justified]{caption}
\usepackage{dcolumn}
\usepackage{bm}
\usepackage{amsmath}
\usepackage{booktabs}
\usepackage{physics}

\begin{document}

\preprint{APS/123-QED}

\title{Efficient optical configurations for trapped-ion entangling gates}

\author{Aditya Milind Kolhatkar}
 \email{Contact author: amk394@cornell.edu}
\author{Karan K. Mehta}

\affiliation{
 School of Electrical and Computer Engineering, Cornell University, Ithaca, NY 14853}

\date{\today}

\begin{abstract}
High-fidelity and parallel realization in scalable platforms of the two-qubit entangling gates fundamental to universal quantum computing constitutes one of the largest challenges in implementing fault-tolerant quantum computation. Integrated optical addressing of trapped ions offers routes to scaling the high-fidelity optical control demonstrated to date in small  systems. Here we show that capabilities practically enabled by integrated optics can additionally alleviate laser powers required for both light-shift (LS) and M{\o}lmer-S{\o}rensen (MS) geometric phase gates acting on long-lived ground-state qubit encodings in a broad range of ion species. We present a theoretical analysis of spontaneous photon scattering (SPS) in stimulated Raman processes driven by spatially structured drive fields, which we employ to assess trapped-ion gates utilizing carrier nulling via ion positioning at phase-stable standing-wave (SW) nodes. Our calculations indicate that suppressed SPS at intensity nodes allows for gate drives operating at smaller Raman detunings and as a result approximately an order-of-magnitude lower power (with significantly larger enhancement in certain parameter regimes) for gates of a given duration and scattering-limited fidelity as compared to gates using running-wave (RW) fields. The SW schemes have the additional benefit of eliminating undesired coherent couplings that can limit gate speeds. Our work quantifies power requirements for multiple ion species and enhancements to be expected from carrier-nulled configurations practically enabled by integrated delivery, and informs experiments and systems for realization of fast and power-efficient laser-based entangling gates in scalable platforms. Our analysis further suggests potential for structured drive fields to mitigate power and scattering limitations in stimulated Raman processes coupling to atomic motion more broadly in both trapped ions and neutral atoms. 
\end{abstract}

\maketitle
\section{Introduction}

Entangling two-qubit gates are generally the most challenging operations to apply with high fidelities in quantum information systems. While high-fidelity gates have been demonstrated in trapped-ion systems \cite{bruzewicz2019trapped} with both laser \cite{ballance2016high, gaebler2016high, clark2021high, moses2023race} and microwave \cite{ospelkaus2011microwave, harty2016high, zarantonello2019robust, srinivas2021high, loschnauer2024scalable, nunnerich2025fast} gate drives, further improvements in fidelity and gate rates will be critical for minimizing resource overheads associated with quantum error correction \cite{steane2003overhead} and for robustness in large-scale platforms. Both in current demonstrations and with respect to future improvements, the high optical or microwave powers required present a key challenge. For long-lived Zeeman or hyperfine qubit encodings \cite{langer2005long, ruster2016long, wang2021single}, intensities required for laser-driven gates of given duration and fidelity are fundamentally set by the spontaneous photon scattering (SPS) accompanying stimulated Raman dynamics, analyzed in previous work for plane-wave drive fields \cite{ozeri2007errors, uys2010decoherence, sawyer2021wavelength, moore2023photon}. The resulting power requirements for high-fidelity gates pose critical architectural challenges for large-scale systems \cite{de2021materials, brown2021materials, moody20222022}; as a result, reducing power requirements for entangling logic is a challenge of both fundamental and practical importance, and will likely be critical to enabling further improvements in quantum logic performance in large-scale platforms.

Waveguide-based optical delivery within ion trap devices \cite{mehta2016integrated, niffenegger2020integrated, mehta2020integrated, ivory2021integrated, moody20222022, vasquez2023control, clements2026sub} may address challenges in scaling optical control across parallel zones \cite{kwon2024multi, mordini2024multi}. The passive optical phase and amplitude stability afforded by such integration also enables delivery of optical field profiles in which the atom-light interaction can be tailored to alleviate bottlenecks on basic physical operations \cite{mehta2019towards, peshkov2023excitation, vasquez2023control, clements2026sub, xing2025trapped} without need for active pointing or path-length stabilization \cite{schmiegelow2016phase, saner2023breaking}. 

\begin{figure*}[t]
\centering
\includegraphics[width=0.65\linewidth]{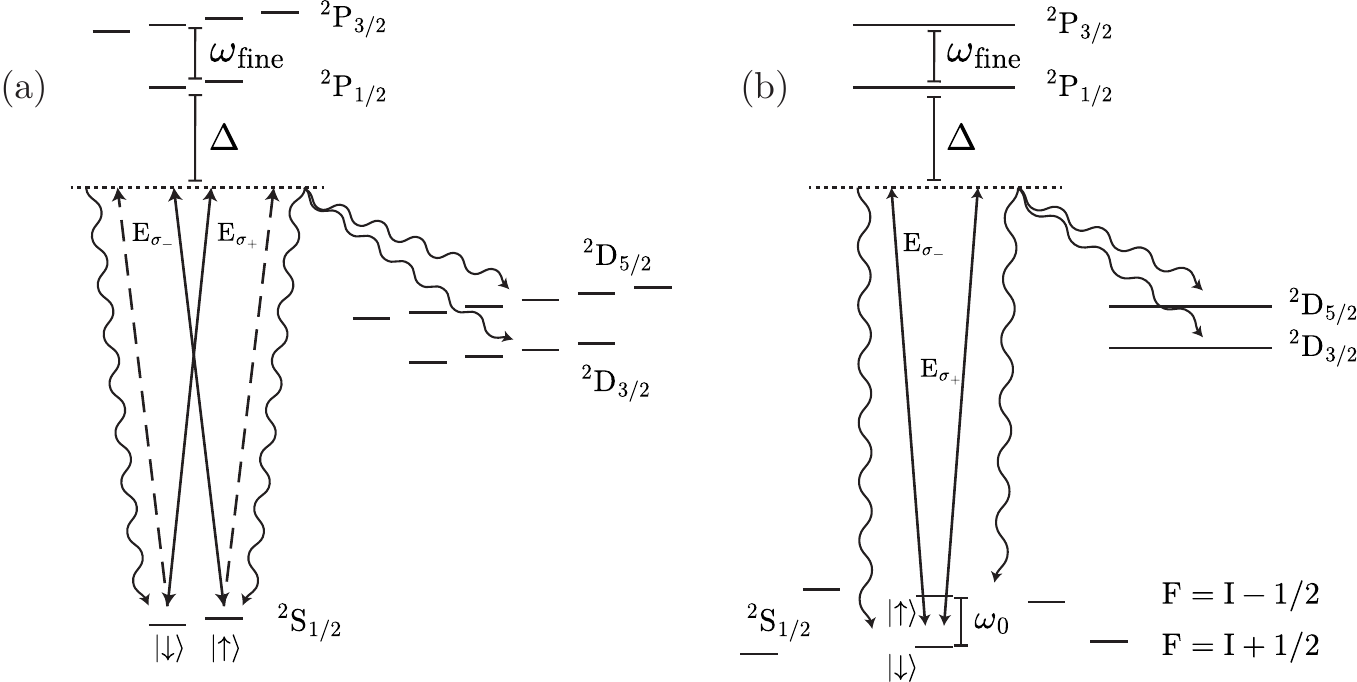}
\caption{Schematic level structures of the qubit encodings considered here. (a) We consider light shift (LS) gates performed on qubits encoded in the Zeeman sublevels of zero-nuclear-spin ($I=0$) species, $\ket{\downarrow}=\ket{{}^{2}S_{1/2}; m_J=-1/2
}$ and $\ket{\uparrow}=\ket{{}^{2}S_{1/2}; m_J=+1/2}$. Straight lines with arrows illustrate coupling of the qubit states to the sublevels in the $P$ manifolds via far-detuned Raman fields. Arrows with dashes indicate that the state $\ket{\downarrow}$ couples more weakly to the circular polarization component $E_{\sigma_-}$ of the Raman fields (since it couples via this component only to the ${}^2\mathrm{P}_{3/2}$ manifold), and vice versa for the state $\ket{\uparrow}$. Wavy lines indicate sponetaneous photon scattering (SPS) events. (b) We also consider M\o lmer-S\o rensen (MS) gates performed on qubits encoded in the ``clock'' transition, $\ket{\downarrow}=\ket{{}^{2}S_{1/2};F=I+1/2,m_{F}=0}$ and $\ket{\uparrow}=\ket{{}^{2}S_{1/2};F=I-1/2,m_{F}=0}$, in species with $I\neq 0$ \cite{ozeri2007errors}. In both cases, a small static magnetic field lifts the degeneracy of the Zeeman sublevels.}
\label{fig:level_struct}
\end{figure*}

Here, we propose a scheme for implementing geometric phase gates on ground state qubit encodings with stimulated Raman transitions driven using a combination of a phase-stable standing wave (SW) and a running wave (RW). Key to our analysis of the SW-based scheme is a detailed treatment of fundamental gate errors due to the effects of SPS on internal and motional states for drive fields with spatially varying intensity profiles. Due to suppressed SPS for ions positioned at SW intensity nodes, for a given gate drive strength and SPS error, the scheme allows lower Raman detunings and hence required power. The proposed scheme additionally suppresses off-resonant ``carrier" or ``direct-drive" coupling in the Hamiltonian \cite{sorensen2000entanglement, langer2006high} that can limit gate speeds achievable \cite{steane2014pulsed, schafer2018fast, ballance2016high}, as has been explored and demonstrated for M{\o}lmer-S{\o}rensen (MS) gates driven on optical qubits \cite{mehta2019towards, saner2023breaking}.

We show that entangling gates of a given duration and fidelity can be implemented with approximately an order of magnitude lower total optical power relative to conventional RW-based implementations. Due to the scaling of SPS rates with detuning, the power enhancement can be significantly larger in certain parameter regimes depending on ion species, particularly at large detunings and high fidelities. Alternatively, for a fixed total power, the scheme allows $>10\times$ faster gates, which can significantly reduce effects of miscalibrations and drifts whose impacts often scale quadratically with the gate time $\tau_{\mathrm{g}}$, as well as infidelities due to electric field noise and motional mode heating that scale linearly with $\tau_g$ \cite{ballance2017high}. 

\begin{figure*}[t]
\centering
\includegraphics[width=\linewidth]{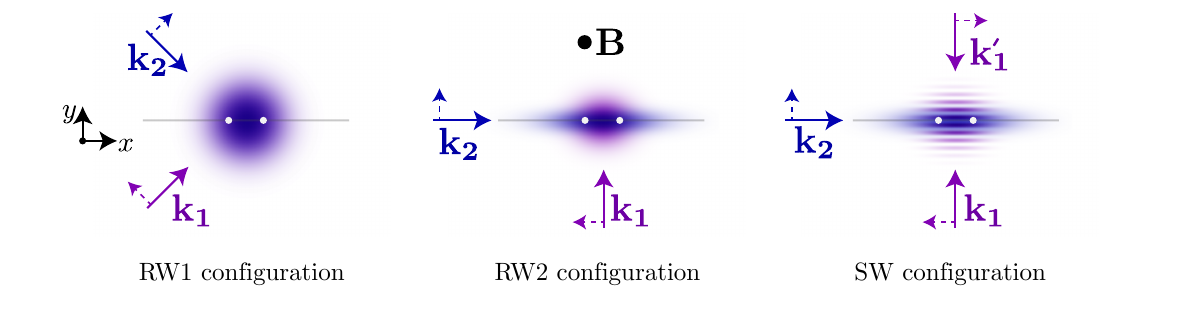}
\caption{Beam configurations for two-qubit gates on a radial mode using integrated optics, for qubits encoded in the ground state manifold. We assume that all beams are sourced from quasi-TE waveguide modes and are linearly polarized in the plane of the page (dashed arrows). Wave-vectors labeled have components in-plane ($k_\parallel$) and along the $z$-direction ($k_\perp$), with only the in-plane components drawn. The gray line denotes the trap axis with white dots showing the ion locations. (Left) We consider, as the most direct application of the conventional gate drive, a beam configuration that employs a pair of running wave (RW) Raman fields. The wave-vector difference $\mathbf{k_1} - \mathbf{k_2}$ lies along $\mathbf{u_y}$ maximize the drive strength for this radial direction. (Center) RW2 is a variation of the RW1 configuration where the Raman beams propagate along and transverse to the trap axis, in the plane of the page. This geometry allows for simple use of elliptical beam spots focused tightly along the radial direction to reduce power requirements, at the cost of reducing the coupling to the radial modes and introducing an unwanted coupling to the axial modes. (Right) We introduce a beam configuration where the gate is driven by a combination of a standing wave (SW) and a RW. By placing the ions at a SW intensity null, we implement a carrier-free drive that suppresses gate errors due to spontaneous photon scattering (SPS) relative to the RW configurations. In RW2 and SW different beam waists are depicted for fields 1 and 2 only for illustration; we take equal waists in our analysis. }
\label{fig:beam_geometries_schematic}
\end{figure*}

The predicted enhancements hold for multiple ion species for both typical geometric phase gate implementations, i.e. light-shift (LS) gates and MS gates \cite{sorensen1999quantum, leibfried2003experimental, lee2005phase}, across a range of SPS-limited gate fidelities. By reducing laser powers required to achieve low SPS errors, our scheme may allow laser-based gates to come significantly closer to performance promised by microwave-based quantum logic \cite{ospelkaus2011microwave, harty2016high, zarantonello2019robust, srinivas2021high, loschnauer2024scalable, nunnerich2025fast} which is free of photon-scattering errors, while preserving the attractive features of laser-based gates with respect to addressing and potentially power and thermal considerations. Also preserved is the favorable scaling to faster operation: gates driven through two-photon stimulated-Raman transitions have $\tau_{\mathrm{g}}$ that scales inversely with the peak drive power $P$ rather than $\sqrt{P}$ as for single-photon optical-transition MS \cite{benhelm2008towards, mehta2020integrated} or microwave-driven gates. 

We hence expect the concepts and calculations presented here to inform near-term experiments in integrated platforms, and to play a key role in high-fidelity parallelized laser gate implementations in large-scale trapped-ion processors. Our analysis of SPS at SW intensity nulls also serves as a starting point for the study of more general stimulated-Raman processes in structured light fields in both trapped-ion and neutral atom platforms where, for instance, analogous optical configurations can be used to enhance Raman sideband cooling \cite{kaufman2012cooling, thompson2013coherence, jenkins2022ytterbium} and higher-order coherent interactions, e.g. state-dependent motional squeezing \cite{katz2022n, buazuavan2024squeezing}. 

Below we present the intuition behind the proposed SW-based gate scheme, considering LS gates implemented on Zeeman sublevels of ions with zero nuclear spin, and for MS gates applied to zero-field clock qubits in species with hyperfine structure; and then in Section \ref{Sec:scatter} outline the methods taken to calculate errors from Raman and Rayleigh SPS accounting for both internal-state and motional decoherence.  We then present results of calculations for the total SPS error in each of these configurations for optimal gate parameters given a fixed gate duration and total available laser power, quantifying the predicted enhancements relative to conventional configurations using similar beam profiles. We conclude with a discussion of additional features and challenges anticipated in experimental implementation. 

\section{Entangling gate configurations for ground-state qubits} \label{Sec:coherent-drive}

\subsection{Qubit encodings} \label{SubSec:qubit}
Given the large number of potential encodings, we focus our discussion and calculations on particular qubit encodings of select ion species. 

For LS gates, we focus on Zeeman qubit encodings in $S_{1/2}$ electron spin states of $I=0$ species (Fig.~\ref{fig:level_struct}a) \cite{kaufmann2017scalable}, with $\ket{\downarrow}=\ket{{}^{2}S_{1/2}; m_J=-1/2
}$ and $\ket{\uparrow}=\ket{{}^{2}S_{1/2}; m_J=+1/2}$, motivated by the potential long-term interest in these simple encodings due to the elimination of qubit leakage pathways \cite{brown2018comparing, brown2019handling}, and since LS gates cannot be directly applied to first-order magnetic field-insensitive ``clock" qubits \cite{lee2005phase}. Here we present results for ${}^{40}\mathrm{Ca}^+$, ${}^{88}\mathrm{Sr}^+$ and ${}^{138}\mathrm{Ba}^+$. A small static magnetic field lifts the degeneracy of the Zeeman sublevels and sets the quantization axis. For MS gates, we consider low-field, approximately field-insensitive ``clock'' states of species with $I\ne0$ employed in many current experiments \cite{debnath2016demonstration, hucul2017spectroscopy, moses2023race} (Fig.~\ref{fig:level_struct}b), with $\ket{\downarrow}=\ket{{}^{2}S_{1/2};F=I+1/2,m_{F}=0}$ and $\ket{\uparrow}=\ket{{}^{2}S_{1/2};F=I-1/2,m_{F}=0}$. Results are presented for ${}^{43}\mathrm{Ca}^+$, ${}^{137}\mathrm{Ba}^+$ and ${}^{171}\mathrm{Yb}^+$ in the main text and for ${}^{9}\mathrm{Be}^+$, ${}^{25}\mathrm{Mg}^+$ and ${}^{87}\mathrm{Sr}^+$ in Appendix \ref{App:scatter}.

While we focus on these qubit encodings for concreteness, the analysis and advantages discussed below can be straightforwardly extended to other qubit encodings, e.g. the ``stretch'' qubit in $I\neq 0$ species or finite-field clock states. 

\subsection{Beam configurations and gate interaction} \label{SubSec:geometry-and-Ham}

Trapped-ion entangling gates are typically implemented as geometric phase gates \cite{sorensen1999quantum, leibfried2003experimental}. Since the mechanisms behind these gates have been discussed extensively in the literature \cite{lee2005phase, ballance2016high}, we only review the principal aspects of the LS interaction here and the mechanism and intuition behind the scheme proposed in this work.  Detailed derivations of the gate Hamiltonians for both LS and MS gates for the field configurations in this paper are presented in Appendix \ref{App:gateham}. 

In a geometric phase gate, a state-dependent force is used to drive motion. In the LS gate, a different force is applied on ions in different eigenstates of $\hat\sigma_z$, whereas in the MS gate, the force is distinct on different eigenstates of a superposition of $\hat\sigma_x$ and $\hat\sigma_y$. Each collective internal state denoted by the double index $s_1, s_2$ becomes entangled with the the motional state during the gate, and additionally acquires a state-dependent geometric phase $e^{i\Phi_{s_1s_2}(t)}$. The phase of the force $F_{s_1s_2}(t)$ is modulated such that at the end of the gate, the motion returns to its initial state and decouples from the internal state. Through careful choice of the gate parameters, the imprinted phases are set such that the operation implements a maximally-entangling phase gate on the two-ion internal state \cite{lee2005phase}.

In Figure \ref{fig:beam_geometries_schematic}, we show beam configurations for three possible implementations relevant to both LS and MS gates acting on ground-state qubits acting on a radial mode of motion. We assume throughout this paper that the beams are sourced from quasi-TE waveguide modes and are oriented in a crossed-linear configuration, where the beam polarizations are perpendicular to each other and to the magnetic field that sets the quantization axis. We also assume that all beams are emitted at an angle $\theta_z$ to the vertical.

Consider first as the most direct application of conventional gate drives the RW1 configuration and the intensity profile of each polarization component therein. Expressed approximately as plane waves near the ion locations, the fields in the two beams can be written as,
\begin{eqnarray}
\mathbf{E_1^\mathrm{RW1}} &=& E_1\left(\frac{\mathbf{-u_x} + \mathbf{u_y}}{\sqrt 2}\right)e^{i\mathbf{k_1}\cdot\mathbf{r}}e^{-i\omega_{\mathrm{las}}t}, \\
\mathbf{E_2^\mathrm{RW1}} &=& E_2\left(\frac{\mathbf{u_x} + \mathbf{u_y}}{\sqrt 2}\right)e^{i\mathbf{k_2}\cdot \mathbf{r}}e^{-i(\omega_{\mathrm{las}}+\Delta\omega)t},
\end{eqnarray} 
and we assume $E_1$ and $E_2$ real for simplicity. Defining the total electric field $\mathbf{E_t} \equiv \mathbf{E_1} + \mathbf E_2$ and the circular polarization unit vectors $\mathbf{u_{\sigma_\pm}} \equiv \frac{\mathbf{u_x} \pm i\mathbf{u_y}}{\sqrt 2}$ given our choice of quantizing $B$-field orientation along $\mathbf{u_z}$, we can write for the intensity in either polarization component:  
\begin{align}
I_{\sigma_\pm} \propto \left| \mathbf{E_t^\mathrm{RW1}} \cdot \mathbf{u_{\sigma_\pm}^*} \right|^2 &= \frac{E_1^2 + E_2^2}{2} \pm E_1E_2\sin\left(\Delta\mathbf{k}\cdot\mathbf{r} - \Delta\omega t\right) \nonumber\\
&= \frac{E_1^2 + E_2^2}{2} \pm E_1E_2\sin\left(\sqrt{2}k_{||}y - \Delta\omega t\right),
\end{align}
where $\Delta \mathbf{k}\equiv\mathbf{k_1}-\mathbf{k_2}$ and  $k_\parallel \equiv k_{\mathrm{las}} \sin\theta_z$ is the in-plane wavevector component. 

With respect to the coherent atom-light coupling, the first term results in a static shift of the energy levels common to both levels for the linearly polarized beams considered here, with negligible impact on dynamics \cite{lee2005phase}. For LS gates, $\Delta\omega$ is chosen close to the driven motional mode frequency, and the spatial gradient of the second term is proportional to the force on the ion. The state-dependence of the force arises from the differential strength of coupling of the polarization components to the qubit levels (Fig.~\ref{fig:level_struct}a); the $\mathbf{\sigma_+}$ component of the field, for instance couples more strongly to the state $\ket{\downarrow}=\ket{S_{1/2},m=-1/2}$ and vice versa. The strength of the differential drive on the two levels is maximum when the beams are arranged in the crossed-linear polarization configuration considered here \footnote{Any $\pi$ polarization component in the beams couples equally to the two levels and does not contribute to the differential drive, so we only consider beams with polarizations perpendicular to the magnetic field. Without loss of generality, we take the polarization vectors to be $\mathbf{u_1}=\sqrt{p_1}\mathbf{u_{\sigma_+}}+\sqrt{1-p_1}\mathbf{u_{\sigma_-}}$ and $\mathbf{u_2}=\sqrt{p_2}e^{-i2\varphi}\mathbf{u_{\sigma_+}}+\sqrt{1-p_2}\mathbf{u_{\sigma_-}}$. The differential force on the two levels can then be written in the form $(\sqrt{p_1(1-p_2)}-\sqrt{p_2(1-p_1)}e^{i2\varphi})(\Omega^{(P_{1/2})}-\Omega^{(P_{3/2})})$, where $\Omega^{(P_{1/2},P_{3/2})}$ denote couplings to the two $P$ manifolds. The differential force can be maximized by picking $p_1=p_2=1/2,\varphi=\pi/2$, in which case the force on the two levels is equal and opposite.}. The force oscillates at the difference frequency $\Delta\omega$, thus driving the motional mode of interest along $\mathbf{u_y}$ near-resonantly. This also allows us to neglect the effect of the drive on the other far-detuned mode(s) that couple to the Stark shift, when the gate dynamics are slow compared to the timescale set by the motional modes (rotating wave approximation). 

The RW2 configuration (Fig.~\ref{fig:beam_geometries_schematic}, center) is similar but considered because it permits simple beam spots with tight focuses to address both ions with higher $E_1, E_2$ for fixed beam powers. However, from the total field in this configuration,
\begin{equation}
\left| \mathbf{E_t^\mathrm{RW2}} \cdot \mathbf{u_{\sigma_\pm}^*} \right|^2 = \frac{E_1^2 + E_2^2}{2} \pm E_1E_2\sin\left(k_{||}y + k_{||}x - \Delta\omega t\right),
\end{equation}
we see that in comparison to the RW1 configuration, the intensity gradient in the $y$-direction is weaker by a factor of $1/\sqrt2$ since $\mathbf{\Delta k}$ lies at $45\degree$ to the trap axis. This geometry results in first-order gradients along the trap axis that produces undesirable couplings to the axial modes. The SW field configuration of particular interest here uses the same elliptical focal spots, allowing direct comparison to this RW equivalent. 

In both RW configurations, besides the intensity gradients responsible for the state-dependent force, the ions experience a static intensity $\propto E_1^2 + E_2^2$ that contributes nothing to gate dynamics while resulting in SPS and the associated decoherence and infidelities \cite{ozeri2007errors}. 

As a means to ameliorate the SPS rate while maintaining the intensity gradient driving the gate, we propose replacing the beam along $\mathbf{k_1}$ in RW2 with a pair of beams with wavevectors $\mathbf{k_1}$ and $\mathbf{k_1'}$ and opposing in-plane components $\pm \mathbf{k_{1,\parallel}}$ forming a standing wave (SW) along the $y$-axis, with intensity nodes running parallel to the $x$- and $z$-axes (Fig.~\ref{fig:beam_geometries_schematic}, right). Normalizing such that a SW field with amplitude $E_1$ is formed by the same total power as the RW fields of the same $E_1$, we write,
\begin{eqnarray}
\mathbf{E_1^\mathrm{SW}} &=& \mathbf{u_x} \sqrt{2} E_1 \sin(k_\parallel y) e^{ik_{\perp}z} e^{-i\omega_{\mathrm{las}} t}, \\ 
\mathbf{E_2^\mathrm{SW}} &=& \mathbf{u_y} E_2 e^{ik_\parallel x + ik_{\perp}z} e^{- i (\omega_{\mathrm{las}} + \Delta\omega) t},
\end{eqnarray}
which results in the following intensity profile: 
\begin{multline}
\left| \mathbf{E_t^\mathrm{SW}} \cdot \mathbf{u_{\sigma_\pm}^*} \right|^2 = \frac12\left(2 E_1^2 \sin^2(k_\parallel y) + E_2^2  \right) \\
\pm \sqrt{2} E_1E_2 \sin(k_\parallel y)\sin(k_\parallel x - \Delta\omega t).
\end{multline}
The gradient along $y$ here has exactly the same strength as in RW1; however the average intensity seen from field $\mathbf{E_1}$ is 0 for ions at $y=0$. The amplitude $E_1$ can be increased and $E_2$ proportionally decreased to keep the first-order gradient and state-dependent force constant, while reducing the SPS rate. The static term associated with $\mathbf{E_1}$ now results in a state-independent shift of the trap frequency owing to its quadratic spatial dependence around $y=0$; and as we discuss in Section \ref{Sec:scatter}, the rate of SPS from $\mathbf{E_1}$ is set by the intensity around $y=0$ sampled by the ion wavepackets during the gate. This determines the optimal distribution of power between the fields $\mathbf{E_1}$ and $\mathbf{E_2}$ and the concomitant reduction in gate errors from SPS in this scheme. These simple considerations indicate qualitatively why the SW configuration should allow reduced SPS scatter for a given total optical power and LS gate drive strength. 

More formally, these beam configurations drive stimulated Raman transitions in the ground state manifold \cite{wineland2003quantum}. For a gate driving the stretch mode $y_s$, the LS Hamiltonian in the interaction picture with respect to the bare Hamiltonian of the internal and motional degrees of freedom of the ions, with the rotating-wave approximation (RWA) applied on terms oscillating at typical trap frequency scales, takes the following form in all three beam geometries (see Appendix \ref{App:gateham}):
\begin{equation}
\hat H^{\mathrm{(LS)}}_{\mathrm{int}}=\hbar\eta\Omega_{\downarrow\downarrow}(\hat\sigma_{z,1}-\hat\sigma_{z,2})\,\hat a e^{i\delta t+i\phi_{\mathrm{m}}}+h.c. \label{eq:H-LS-main-text}
\end{equation}
where $\delta=\Delta\omega-\omega_{y_s}$ is the gate detuning, $\Omega_{\downarrow\downarrow}$ is the AC Stark shift on the state $\ket{0}$ due to the oscillating terms in the intensity profile, $\hat a$ ($\hat a^\dagger$) is the annihilation (creation) operator for mode $y_s$ and $\phi_m$ sets the phase of the force on the motional states. Here we have used the fact that $\Omega_{\downarrow\downarrow}=-\Omega_{\uparrow\uparrow}$ for crossed-linear beam polarizations. The strength of the gate drive in each beam configuration is captured within the Lamb-Dicke parameter $\eta$. Assuming that the two-ion radial modes are rotated such that they are oriented along the $y$- and $z$-axes, we define for RW1 and RW2, $\eta = \mathbf{\Delta k}\cdot\mathbf{u_y}\frac{y_{\mathrm{s}}^{(0)}}{\sqrt{2}}$ and for the SW configuration, $\eta = \sqrt{2}\mathbf{k_1}\cdot\mathbf{u_y}\frac{y_{\mathrm{s}}^{(0)}}{\sqrt{2}}$. Here $y^{(0)}_{\mathrm{s}}=\sqrt{\hbar/2m\omega_{y_s}}$ is the RMS extent of a single-ion ground-state wavefunction at the stretch mode frequency $\omega_{y_s}$, and the additional factor of $\sqrt{2}$ in the SW configuration definitions accounts for the gradient strength for fixed $E_1E_2$.  From the geometry in Fig.~\ref{fig:beam_geometries_schematic}, we see that, 
\begin{align}
\eta^{\mathrm{RW1}} &\equiv k_{\mathrm{las}}y^{(0)}_s\sin\theta_z, \label{eq:eta_RW1} \\
\eta^{\mathrm{RW2}} &\equiv \frac{1}{\sqrt{2}}k_{\mathrm{las}}y^{(0)}_s\sin\theta_z, \label{eq:eta_RW2}\\
\eta^{\mathrm{SW}} &\equiv k_{\mathrm{las}}y^{(0)}_s\sin\theta_z. \label{eq:eta_SW} 
\end{align}
The AC Stark shift in \eqref{eq:H-LS-main-text} can be calculated from the dipole matrix elements as,
\begin{align}
\Omega_{\downarrow\downarrow}=g_1g_2\sum_k\Bigg(&\frac{\bra{\downarrow}\hat{\mathbf{r}}\mathbf{_{el}}\cdot\boldsymbol{\epsilon}_1\ket{k}\bra{k}\hat{\mathbf{r}}\mathbf{_{el}}\cdot\boldsymbol{\epsilon}_2\ket{\downarrow}}{\mu^2(\omega_{kg}-\omega_{\mathrm{las}})} \nonumber\\
&\qquad+\frac{\bra{\downarrow}\hat{\mathbf{r}}\mathbf{_{el}}\cdot\boldsymbol{\epsilon}_2\ket{k}\bra{k}\hat{\mathbf{r}}\mathbf{_{el}}\cdot\boldsymbol{\epsilon}_1\ket{\downarrow}}{\mu^2(\omega_{kg}+\omega_{\mathrm{las}})}\Bigg), \label{eq:two-photon-Rabi-rate-SS}
\end{align}
where the sum is over sub-levels of the $P_{1/2}$ and $P_{3/2}$ manifolds, $\boldsymbol{\epsilon_i}$ is the unit polarization vector of the field $\mathbf{E_i}$, $\omega_{kg}$ is the mean angular frequency of the transition between the intermediate state $k$ and the Zeeman sublevels of the $S_{1/2}$ ground manifold, $g_i\equiv eE_i\mu/2\hbar$, and $\mu$ is the largest dipole matrix element connecting the qubit levels to the manifold of excited states,
\begin{align*}
\mu=&|\bra{J'=3/2, m'_J=3/2} \\
&\qquad\qquad\mathbf{\hat r_{el}\cdot \mathbf{u_{\sigma_+}}}\ket{J=1/2,m_J=1/2}|.
\end{align*}
For the MS gate, we consider a `three-beam configuration', where we send a tone at $\omega_1=\omega_{\mathrm{las}}$ in field $\mathbf{E_1}$, and two tones of equal amplitude at $\omega_{2,\mathrm{b}}=\omega_{\mathrm{las}}+\omega_0+\omega_{y_s}+\delta$ and $\omega_{2,\mathrm{r}}=\omega_{\mathrm{las}}-\omega_0+\omega_{y_s}+\delta$ in field $\mathbf{E_2}$, with $\omega_0$ the splitting between the qubit levels \cite{halijan2005phase}.  

The intensity profiles above offer some insight into the resulting dynamics, for all three beam configurations. As before, the first term in the intensity profile results in a static shift in the energy levels and in the SW configuration a shift in the trap frequency. The second term $\propto E_1E_2$ now oscillates close to frequencies $\omega_0\pm\omega_{y_s}$ of the blue and red motional sidebands of the spin-flip transition. This term therefore stimulates transitions between the qubit levels besides providing the force on the ions, with different eigenstates of the spin-flip operator driven out of phase relative to each other. We show in Appendix \ref{App:gateham} that the MS gate Hamiltonian in the interaction picture and rotating-wave approximation (RWA) with respect to terms oscillating near and above $\omega_{y_s}$ takes the form,
\begin{equation}
\hat H^{\mathrm{(MS)}}_{\mathrm{int}}=\frac{1}{\sqrt{2}}\hbar\eta\Omega_{\downarrow\uparrow}(\hat\sigma_{\phi_s,1}-\hat\sigma_{\phi_s,2})\,\hat a e^{i\delta t+i\phi_{\mathrm{m}}}+h.c., \label{eq:H-MS-main-text}
\end{equation}
where the spin-flip operator $\hat\sigma_{\phi_s,j}=\cos\phi_s\hat\sigma_{x,j}+\sin\phi_s\hat\sigma_{y,j}$ effects internal state transitions and the `spin' phase $\phi_s$ sets the basis of rotation of these transitions in the $xy$-plane of the Bloch sphere. The spin-flip Rabi frequency $\Omega_{\downarrow\uparrow}=\Omega_{\uparrow\downarrow}$ is maximized in the crossed-linear polarization configuration \footnote{Selection rules for $\pi$ polarization prevent transitions between one of the qubit levels and any given sublevel in the $P$ manifold, so only consider polarizations perpendicular to the magnetic field. The coupling between the qubit levels is of opposite sign due to $\sigma_{\pm}$ polarizations, i.e. for polarization vectors $\mathbf{u_1}=\sqrt{p_1}\mathbf{u_{\sigma_+}}+\sqrt{1-p_1}\mathbf{u_{\sigma_-}}$ and $\mathbf{u_2}=\sqrt{p_2}e^{-i2\varphi}\mathbf{u_{\sigma_+}}+\sqrt{1-p_2}\mathbf{u_{\sigma_-}}$, $\Omega_{\downarrow\uparrow}\propto \sqrt{p_1(1-p_2)}-\sqrt{p_2(1-p_1)}e^{i2\varphi}$, which is maximum for crossed linear polarizations.}, and the Lamb-Dicke parameters $\eta$ in the three beam configurations are as defined in eqs.~\eqref{eq:eta_RW1}-\eqref{eq:eta_SW}. The factor of $1/\sqrt{2}$ accounts for the distribution of power between the two tones in $\mathbf{E_2}$.

The spin-flip Rabi frequency in \eqref{eq:H-MS-main-text} can be calculated as,
\begin{align}
\Omega_{\downarrow\uparrow}=g_1g_2\sum_k\Bigg(&\frac{\bra{\downarrow}\hat{\mathbf{r}}\mathbf{_{el}}\cdot\boldsymbol{\epsilon}_1\ket{k}\bra{k}\hat{\mathbf{r}}\mathbf{_{el}}\cdot\boldsymbol{\epsilon}_2\ket{\uparrow}}{\mu^2(\omega_{kg}-\omega_{\mathrm{las}})} \nonumber\\
&\qquad+\frac{\bra{\downarrow}\hat{\mathbf{r}}\mathbf{_{el}}\cdot\boldsymbol{\epsilon}_2\ket{k}\bra{k}\hat{\mathbf{r}}\mathbf{_{el}}\cdot\boldsymbol{\epsilon}_1\ket{\uparrow}}{\mu^2(\omega_{kg}+\omega_{\mathrm{las}})}\Bigg), \label{eq:two-photon-Rabi-rate-spin-flip}
\end{align} 
where the largest dipole matrix element $\mu$ is,
\begin{align*}
\mu=&|\bra{F'=I+3/2, m'_F=I+3/2} \\
&\qquad\qquad\mathbf{\hat r_{el}\cdot \mathbf{u_{\sigma_+}}}\ket{F=I+1/2,m_F=I+1/2}|.
\end{align*}
We use the `phase-insensitive' frequency configuration \cite{halijan2005phase} in our derivation of the Hamiltonian in Appendix~\ref{App:gateham} as a particular example; our results for the gate error due to spontaneous photon scattering in Section~\ref{Sec:scatter} remain valid for the `phase-sensitive' frequency configuration, where the spin phase becomes sensitive to path-length fluctuations between the two Raman beams. 

Besides suppressing SPS in the same manner in the MS gate as described for LS gates above, the SW scheme implements precisely the kind of carrier-nulled drive previously considered for optical qubits \cite{mehta2019towards, saner2023breaking}, owing to the zero-crossing of $\mathbf{E_2}$ at $y=0$ (see Appendix \ref{App:gateham}). Note that by sending only a single tone into the SW field profile $\mathbf{E_1}$, this configuration avoids oscillating perturbations to the trap frequency and potential associated undesirable squeezing effects \cite{vasquez2024state}.

During the gate, the ions undergo coherent displacements $\ket{\alpha^{}_{s_1s_2}(t)}$ in the $x-p$ phase space conditioned on the two-qubit internal state. Here, the basis states are $s_1,s_2\in\{\downarrow,\uparrow\}$ for LS gates and $s_1,s_2\in\{+_{\phi_s},-_{\phi_s}\}$ for MS gates, where $\ket{\pm_{\phi_s}}$ are the eigenstates of $\hat\sigma_{\phi_s}$. The motion is unentangled from the spin at the end of the gate. For a fixed gate time $\tau_{\mathrm{g}}$, in a $K$-loop maximally-entangling gate, the gate detuning is set to $\delta=2\pi K/\tau_{\mathrm{g}}$ and the gate drive obeys the following constraint \cite{langer2006high, lee2005phase}:
\begin{equation}
\eta\,\Omega_{\mathrm{gate-drive}}\tau_{\mathrm{g}}=\pi\sqrt{K}, \label{eq:Omega-gate-drive-constraint}
\end{equation}
where $\Omega_{\mathrm{gate-drive}}=2|\Omega_{\downarrow\downarrow}|$ for the LS gate and $\Omega_{\mathrm{gate-drive}}=\sqrt{2}|\Omega_{\downarrow\uparrow}|$ for the MS gate.

\section{Gate error due to spontaneous photon scattering} \label{Sec:scatter}
We proceed to calculate the rates of SPS and associated effects on both internal and motional coherence during phase gates for the beam configurations of Fig.~\ref{fig:beam_geometries_schematic}. 

The trade-off between gate error and laser power has been previously studied for gates with conventional RW-based drives \cite{ozeri2007errors, uys2010decoherence, sawyer2021wavelength, moore2023photon, carter2023}. In this section we consider spontaneous photon scattering due to a SW field and show that the SPS rate due to the SW is set by the rate of absorption on the red and blue motional sidebands, and therefore, suppressed on the order of $\eta^2$ relative to a RW field of the same waists and total power. Unlike absorption on the carrier, which determines the SPS rate due to a RW \cite{ozeri2007errors}, the rate of absorption on the sidebands depends on the collective motional state, so we integrate the instantaneous scattering rate over the motional trajectory to calculate the probability of photon scattering during the gate. We also model the impact on motional coherence of sideband excitations induced during SPS events from the SW field as damping due to an infinite-temperature bath \cite{turchette2000motional}, which in contrast to the internal-state decoherence is not suppressed for the SW field relative to RW fields. Finally, we quantify the advantage offered by the SW-based gate drive by comparing SPS-induced gate errors in the SW configuration with those in RW1 and RW2.

Consider a single ion with the qubit encoded in the ground state manifold illuminated by a laser field detuned from all dipole transitions. To calculate the SPS rate due to this field, we use the Kramers-Heisenberg formula based on second-order perturbation theory \cite{loudon2000quantum} that describes the rate of scattering from atomic state $\ket i$ to $\ket f$: 
\begin{equation}
\Gamma_{if} = \Omega^{2}_{\mathrm{L}}\sum_{\epsilon_{\mathrm{sc}}}\left | \sum_{e} \chi^{(e,\epsilon_{\mathrm{sc}})}_{if}\right |^2, \label{eq:scattering_rate_if}
\end{equation}
where $e$ indexes the possible excited states, and $\epsilon_\mathrm{sc} = 0, \pm 1$ indexes the possible polarizations $\mathbf{u_\pi}$, $\mathbf{u_{\sigma_\pm}}$ for the spontaneously scattered photon. $\Omega_{\mathrm{L}}^2$ is proportional to the excitation rate due to the laser field. In general this depends on the internal state of the qubits, as we will see below in the case of the SW configuration. 

The amplitudes $\chi^{(e,\epsilon_{\mathrm{sc}})}_{if}$ of each scattering path are calculated from the dipole matrix elements by treating each scattering event as a two-photon process: stimulated absorption of the incident light accompanied by a transition from the initial state $\ket{i}$ to an intermediate excited state $\ket{e}$, and spontaneous emission of a photon with polarization indexed by $\epsilon_\mathrm{sc}$ into the vacuum modes of the EM field accompanied by a transition to the final state $\ket{f}$ \cite{ozeri2007errors, moore2023photon}:
\begin{widetext}
\begin{equation}
\chi^{(e,\epsilon_{\mathrm{sc}})}_{if} = \sqrt{A_{J_e,J_i}A_{J_e,J_f}} \sqrt{\frac{(\omega_{\mathrm{las}}-\omega_{fi})^{3}} {\omega_{ei}^3\omega_{ef}^3}}\frac{1}{\mu}\left(\frac{\bra{f}\mathbf{\hat r_{el}}\cdot \mathbf{u_{sc}} \ket{e}\bra{e}\mathbf{\hat r_{el}}\cdot \mathbf{u_{las}}\ket{i}}{\omega_{ei}-\omega_{\mathrm{las}}} +\frac{\bra{f}\mathbf{\hat r_{el}}\cdot \mathbf{u_{las}}\ket{e}\bra{e}\mathbf{\hat r_{el}}\cdot \mathbf{u_{sc}}\ket{i}}{\omega_{ef}+\omega_{\mathrm{las}}}\right). \label{eq:chi}
\end{equation}
\end{widetext}
Here, $\omega_{ab}$ is the angular frequency of the transition between states $\ket a$ and $\ket b$, $A_{J_e,J_i}$ ($A_{J_e,J_f}$) is the rate of spontaneous scattering from the excited state $\ket{e}$ to $\ket{i}$ ($\ket{f}$) and  $\mathbf{u_{las}}$ and $\mathbf{u_{sc}}$ denote the polarization unit vectors of the Raman and spontaneous scattered fields, respectively.

In what follows, the average gate infidelity due to spontaneous photon scattering (SPS) will be calculated as \cite{ozeri2007errors},
\begin{equation}
\epsilon_{\mathrm{SPS}} = \epsilon_{\mathrm{qub}}+\epsilon_{\mathrm{mot}},
\end{equation}
where $\epsilon_{\mathrm{qub}}$ denotes the probability of error due to decoherence of the qubits' internal states during the gate, and $\epsilon_{\mathrm{mot}}$ denotes the error due to motional decoherence. We will now consider the different possible scattering processes and describe how we quantify the errors from each for both the RW and SW fields in the field configurations of Fig.~\ref{fig:beam_geometries_schematic}. 

\subsection{Qubit decoherence due to Raman scattering} \label{SubSec:scatter-Raman}
Photon scattering is classified as Raman scattering when the final state of the atom $\ket{f}$ differs from its initial state $\ket{i}$. The energy and polarization of the scattered photon carry information about the final internal state $\ket{f}$, so all such events destroy spin coherence in the reduced basis of the two-qubit superposition \cite{ozeri2007errors}. For all configurations considered here, we calculate the Raman scattering rate due to beam $k$ by summing eq.~\eqref{eq:scattering_rate_if} over all pairs of states $\ket{f_j}\neq \ket{i_j}$ for the ions $j=1,2$ \footnote{The expression \eqref{eq:chi} for $\chi^{(e,\epsilon_{\mathrm{sc}})}_{if}$ carries a dependence on the Raman beam frequency and polarization. but for the far-detuned linearly polarized beams considered here, the scattering amplitudes due to the two Raman fields in each beam configuration differ at most by an overall phase.}:
\begin{equation}
\Gamma_{\mathrm{Raman},k} = \sum_{i_1,i_2} p_{i_1i_2} \Omega^{2}_{\mathrm{L},k} \sum_{j}\sum_{f_j\neq i_j}\sum_{\hat{\epsilon}_{\mathrm{sc}}}\left |\sum_e\chi^{(e,\epsilon_{\mathrm{sc}})}_{i_jf_j}\right |^2. \label{eq:decoherence_rate_Raman}
\end{equation}
$p_{i_1i_2}$ here is the probability of the two ions being in the internal state $i_1i_2\in\{\downarrow\downarrow,\downarrow\uparrow,\uparrow\downarrow,\uparrow\uparrow\}$ during the gate. We take $p_{i_1i_2}=1/4$ for all pairs $i_1i_2$ to find the average gate infidelity. We calculate the gate error for each configuration by independently summing the rates of scattering $\Gamma_{\mathrm{Raman},k}$ from the two fields $\mathbf{E_k}$ shown in Fig.~\ref{fig:beam_geometries_schematic} and integrating the rate over the gate duration:
\begin{equation}
\epsilon_{\mathrm{qub,Raman}} = \int_0^{\tau_{\mathrm{g}}}dt\,(\Gamma_{\mathrm{Raman},1}+\Gamma_{\mathrm{Raman},2}). \label{eq:err-qub-Raman}
\end{equation}
The expressions for $\Gamma_{\mathrm{Raman},k}$ in the three configurations differ only in the excitation rates $\Omega_{\mathrm{L},k}^2$. In beam configurations RW1 and RW2, $\Omega_{\mathrm{L},k}^2$ is independent of the qubit state, and proportional to the rate of absorption on the carrier \cite{ozeri2007errors, moore2023photon}:
\begin{equation}
\Omega_{\mathrm{L},k}^2=g_k^2=\left(\frac{eE_k\mu}{2\hbar}\right)^2,\quad \text{(RW1, RW2 Schemes)} \label{eq:Omega_L_RW}
\end{equation}
so the $\Gamma_{\mathrm{Raman},k}$ is constant over the gate duration.

Now consider Raman scattering in the SW scheme. When $\mathbf{E_1}$ is a SW with the ion located at an intensity node, in the Lamb-Dicke regime, the only absorption events are due to sideband transitions on motional modes with non-zero projection onto $\mathbf{k_1}$. Therefore, the effective excitation rate $\Omega_{\mathrm{L}}^2$ is obtained by summing the rates of transition on all red sidebands (RSBs) and blue sidebands (BSBs). As we show in Appendix \ref{App:scatter}, for the LS gate, we may write,
\begin{equation}
(\Omega^2_{\mathrm{L},1}){}_{i_1i_2}= \sum_l \left(\Omega^{(l)2}_{\mathrm{RSB},i_1i_2}+\Omega^{(l)2}_{\mathrm{BSB},i_1i_2}\right),
\end{equation}
where our notation reflects that the sideband absorption rates depend on the two-qubit internal state, since the motional trajectories are state-dependent. $l$ here indexes the modes of motion. The transition rates owing to the sideband couplings have been defined as, 
\begin{eqnarray}
\Omega^{(l)2}_{\mathrm{RSB},i_1i_2} &\equiv& \eta_l^2g_1^2\sum_{n_l=0}^\infty \left|\bra{n_l}\hat a_l\ket{\alpha^{(l)}_{i_1i_2}(t)}\right|^2 \\
\Omega^{(l)2}_{\mathrm{BSB},i_1i_2} &\equiv& \eta_l^2g_1^2\sum_{n_l=0}^\infty \left|\bra{n_l}\hat a^\dagger_l\ket{\alpha^{(l)}_{i_1i_2}(t)}\right|^2
\end{eqnarray}
where $\eta_l\equiv(\mathbf{k_1}\cdot\mathbf{u_y})y_{l}^{(0)}$. We see that $\Gamma_{\mathrm{Raman},1}$ in the SW configuration evolves during the gate in accordance with the trajectory of the motional state $\ket{\alpha^{(l)}_{i_1i_2}(t)}$. Summing the RSB and BSB rates, the instantaneous rate of absorption due to the SW is given by
\begin{equation}
(\Omega^{2}_{\mathrm{L},1}(t)){}_{i_1i_2} = g_1^{2}\sum_l\eta_l^2\left(2|\alpha^{(l)}_{i_1i_2}(t)|^2+1\right),\;\,\text{(SW Scheme)} \label{eq:Omega-L-SW}
\end{equation}
and the excitation rate from field $\mathbf{E_2}$ is $\Omega_{\mathrm{L},2}^2=g_2^2$ as in the RW configurations. The average occupancies of the internal states remain uniform in the MS gate, where the state-dependent coherent displacements $\ket{\alpha^{(l)}_{s_1s_2}(t)}$ occur in the qubit $xy$-basis, with $s_1,s_2\in\{+_{\phi_{\mathrm{s}}},-_{\phi_{\mathrm{s}}}\}$, so the expression analogous to eq.~\eqref{eq:Omega-L-SW} applies. In this fashion, the SPS rate per unit power due to the SW is suppressed by the factor on the order of $\eta^2$ relative to a RW field of the same amplitude. 

\subsection{Qubit decoherence due to Rayleigh scattering} \label{SubSec:scatter-Rayleigh}
Spontaneous scattering events where the qubit returns to its initial state, $\ket{f}=\ket{i}$, are classified as Rayleigh scattering events. In Ref. \cite{uys2010decoherence}, it was shown that the rate of qubit decoherence due to Rayleigh scattering is set by the squared difference of the elastic scattering amplitudes $\chi^{(e,\epsilon_{\mathrm{sc}})}_{ii}$ on the two qubit levels. We estimate the associated gate infidelity as,
\begin{equation}
\epsilon_{\mathrm{qub,Rayleigh}} = \int_0^{\tau_{\mathrm{g}}}dt\,(\Gamma_{\mathrm{Rayleigh},1}+\Gamma_{\mathrm{Rayleigh},2}), \label{eq:err-qub-Rayleigh}
\end{equation}
where the decoherence rate due to each beam, averaged over the occupancies of the two-qubit internal states is \cite{sawyer2021wavelength},
\begin{equation}
\Gamma_{\mathrm{Rayleigh}}=\Omega_{\mathrm{L}}^2 \sum_j\frac{1}{2}\sum_{\epsilon_{\mathrm{sc}}}\left |\sum_e\left(\chi^{(e,\epsilon_{\mathrm{sc}})}_{00}-\chi^{(e,\epsilon_{\mathrm{sc}})}_{11}\right)\right |^2, \label{eq:Gamma-Rayleigh}
\end{equation}
with the excitation rates $\Omega_{\mathrm{L}}^2$ in the three configurations the same as those in the case of Raman scattering. We note that in using eq.~\eqref{eq:Omega-L-SW} to calculate the SW Rayleigh decoherence rate in eq.~\eqref{eq:Gamma-Rayleigh}, we neglect potential subtle effects associated with spin-motion entanglement during the gate; a more careful treatment of Rayleigh dephasing in a SW would require us to study the evolution of the joint density matrix of the internal and external states of the two ions \cite{uys2010decoherence}. In Appendix \ref{App:scatter}, however, we show that the advantage in the power requirement in the SW scheme continues to hold for a pessimistic model of Rayleigh decoherence that assumes that all Rayleigh scattering events due to the SW field lead to a gate error.

Summing the error contributions from the Raman and Rayleigh processes, we obtain, $\epsilon_{\mathrm{qub}}=\epsilon_{\mathrm{qub, Raman}}+\epsilon_{\mathrm{qub, Rayleigh}}$.

\subsection{Motional decoherence due to Rayleigh scattering} \label{SubSec:scatter-Rayleigh-motion}
The random exchange of momentum between the ion and the scattered photon during both Raman and Rayleigh events can contribute to gate infidelity. However, since all Raman scattering events lead to gate error, their effect on motion need not be additionally accounted for \cite{ozeri2007errors, moore2023photon}.

Distinct mechanisms govern the motional decoherence induced by scattering from a RW and a SW. For RW fields, the error can be modeled as arising from a random momentum kick from the emitted photon \cite{ozeri2007errors}. If the scattered photon is emitted along the wavevector $$\mathbf{k_{sc}}=k_{\mathrm{las}}\left(\sin\theta_{\mathrm{sc}}\cos\phi_{\mathrm{sc}}\mathbf{u_x}+\sin\theta_{\mathrm{sc}}\sin\phi_{\mathrm{sc}}\mathbf{u_y}+\cos\theta_{\mathrm{sc}}\mathbf{u_z}\right),$$the displacement $\beta$ in phase space of the mode $y_s$ is,
\begin{align}
|\beta| &=\frac{1}{\sqrt{2}}|(\mathbf{k_{las}-\mathbf{k_{sc}})\cdot \mathbf{u_y}}|\,y^{(0)}_s \nonumber\\
&=\frac{1}{\sqrt{2}}|k_{\mathrm{las},y}-k_{\mathrm{las}}\sin\theta_{\mathrm{sc}}\sin\phi_{\mathrm{sc}}|\,y^{(0)}_s.
\end{align}
We have taken $|\mathbf{k_{sc}}|=k_{\mathrm{las}}$ by energy conservation. The average gate error due to the random displacements was calculated in Ref. \cite{ozeri2007errors}:
\begin{align}
\epsilon_{\mathrm{mot}} &= \frac{\tau_g}{2}\left(\langle|\beta_1|\rangle^2g_1^2+\langle|\beta_2|\rangle^2g_2^2\right) \sum_{i,\hat{\epsilon}_{\mathrm{sc}}}\left |\sum_e\chi^{(e,\epsilon_{\mathrm{sc}})}_{ii}\right |^2. \nonumber \\
&\qquad\qquad\qquad\qquad\text{(RW1, RW2 Schemes)} \label{eq:err-mot-RW}
\end{align}
The mean squared displacement is calculated as,$$\langle|\beta|^2\rangle=\frac{1}{4\pi}\int_0^{2\pi}d\phi_{\mathrm{sc}}\int_{0}^\pi d\theta_{\mathrm{sc}}\sin\theta_{\mathrm{sc}}P(\theta_{\mathrm{sc}},\phi_{\mathrm{sc}})|\beta|^2,$$where $P(\theta_{\mathrm{sc}},\phi_{\mathrm{sc}})$ is the angular probability distribution of $\mathbf{k_{sc}}$. Since the Raman beams we consider do not have a polarization component along the magnetic field, the emission pattern of $\mathbf{k_{sc}}$ is that of a $\sigma_{\pm}$-polarized dipole, so we use$$P(\theta_{\mathrm{sc}},\phi_{\mathrm{sc}})=\frac{3}{4}(1+\cos^2\theta_{\mathrm{sc}}).$$For the two conventional configurations, we find that
\begin{align}
\langle|\beta_1|^2\rangle&=\frac{9}{20}\eta^2,\,\langle|\beta_2|^2\rangle=\frac{9}{20}\eta^2,\,\text{(Scheme RW1)} \\
\langle|\beta_1|^2\rangle&=\frac{7}{10}\eta^2,\,\langle|\beta_2|^2\rangle=\frac{1}{5}\eta^2.\,\text{(Scheme RW2)} \label{eq:beta-sq-RW2}
\end{align}

For the SW field, since absorptions occur through sideband transitions, every excitation adds or subtracts a phonon from the motional state. The scattering events are random, so the sideband excitations induce motional decoherence. Since the relative rates of motional excitation and de-excitation due to absorption on the blue and red sidebands $ (\bar n+1) / (\bar n)$ follow the same relative jump rates for  coupling to an infinite-temperature bath \cite{turchette2000motional}, we model the effect of the sideband absorption events on the motion as a heating process with associated Lindblad jump operators $\hat L_+ = \sqrt{\gamma_{\mathrm{mot}}}\hat a^\dagger$ and $\hat L_- = \sqrt{\gamma_{\mathrm{mot}}}\hat a$. The effective heating rate $\gamma_{\mathrm{mot}}$ in this model, 
\begin{equation}
\gamma_{\mathrm{mot}} \equiv \eta^2g_1^2\sum_{i,\hat{\epsilon}_{\mathrm{sc}}}\left |\sum_e\chi^{(e,\epsilon_{\mathrm{sc}})}_{ii}\right |^2
\end{equation}
is the prefactor of the absorption rates on the gate mode's RSB and BSB resulting in Rayleigh scattering, which are $\gamma_\mathrm{mot} \bar n_{y_s}$ and $\gamma_\mathrm{mot}(\bar n_{y_s}+1)$ respectively. The recoil of the emitted photon results in a correction to $\gamma_{\mathrm{mot}}$ of order $\eta^4$, so we neglect it here. 
The resulting single-loop gate error rate is $\gamma_{\mathrm{mot}}/2$ \cite{ballance2016high}, so we have,
\begin{align}
\epsilon_{\mathrm{mot}} &= \frac{\tau_g}{2}\left(\eta^2g_1^2+\frac{1}{5}\eta^2g_2^2\right) \sum_{i,\hat{\epsilon}_{\mathrm{sc}}}\left |\sum_e\chi^{(e,\epsilon_{\mathrm{sc}})}_{ii}\right |^2, \nonumber \\
&\qquad\qquad\qquad\qquad\qquad\qquad\quad\text{(SW Scheme)} \label{eq:err-mot-SW}
\end{align}
where we have used the same $\langle|\beta_2|^2\rangle=\eta^2/5$ found for the RW2 scheme. 

Because absorptions from the SW profile occur entirely through RSB and BSB couplings, the \textit{fraction} of absorption events that add or subtract a quantum of motion is of order $\eta^{-2}$ larger than for a RW field profile. As a result, even though the total absorption rate for a given power is suppressed by a factor of order $\eta^2$, in contrast to decoherence of the internal state, the error due to motional decoherence is not suppressed for the SW profile. 

\subsection{Calculating power requirements} \label{SubSec:scatter-vs-power}
For a given total available laser power $P_{\mathrm{tot}}$ and target gate time $\tau_{\mathrm{g}}$, we calculate the minimum achievable $\epsilon_{\mathrm{SPS}}$ as described in the preceding discussion, for each beam configuration, which includes determining the optimal power distribution between fields $\mathbf{E_1}$ and $\mathbf{E_2}$. 

In configurations RW1 and RW2, distributing power equally between the two fields is optimal since both fields induce the same gate error per unit power \footnote{\label{note:power-dist}This is not strictly true in the case of the RW2 configuration, since the motional decoherence rate from the two Raman beams is different, see eq.~\eqref{eq:beta-sq-RW2}. However, we assume for simplicity that the total power is distributed equally between the Raman beams in both RW2 and RW1.}, and equal distribution maximizes the two-photon drive \eqref{eq:two-photon-Rabi-rate-SS} and \eqref{eq:two-photon-Rabi-rate-spin-flip} given $P_{\mathrm{tot}}$ for LS and MS gates respectively. We obtain the operating optical frequency $\omega_{\mathrm{las}}$ in terms of $P_{\mathrm{tot}}$ and $\tau_{\mathrm{g}}$ using the constraint \eqref{eq:Omega-gate-drive-constraint}. We then calculate the total gate error $\epsilon_\mathrm{SPS}$ by summing the contributions \eqref{eq:err-qub-Raman}, \eqref{eq:err-qub-Rayleigh} and \eqref{eq:err-mot-RW}.

In the SW configuration, the SPS rate per unit power induced by the SW field is suppressed relative to the RW field, so in general, equal distribution of powers between the SW and the RW is not optimal. For a given $P_{\mathrm{tot}}$, let $p_1$ be the fraction of power in the SW field $\mathbf{E_1}$. Then, the operating laser frequency $\omega_{\mathrm{las}}$ is calculated as a function of $p_1$ by inverting the gate drive constraint of eq.~\eqref{eq:Omega-gate-drive-constraint}. We then find the minimum attainable value of the gate error, i.e. the sum of contributions \eqref{eq:err-qub-Raman}, \eqref{eq:err-qub-Rayleigh} and \eqref{eq:err-mot-SW}, by minimizing the resulting function $\epsilon_{\mathrm{SPS}}(p_1)$.  

\begin{figure*}
\centering
\includegraphics[width=\linewidth]{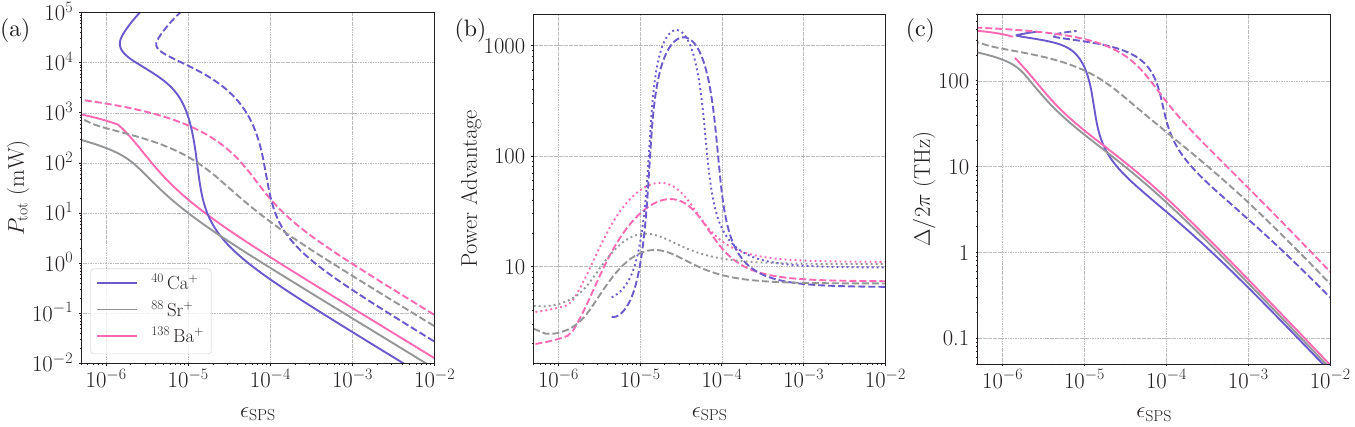}
\caption{Comparison of the laser power requirement for LS gates with gate time $\tau_{\mathrm{g}}=50$ $\mu$s. All beam configurations are in reference to Fig.~\ref{fig:beam_geometries_schematic}. (a) Total laser power required given a target SPS-induced gate error $\epsilon_{\mathrm{SPS}}$, for the SW scheme (solid line) and the RW2 scheme (dashed line). Curves for the RW1 scheme have been omitted for clarity. (b) The ratios of required total laser power $P_{\mathrm{RW2}}/P_{\mathrm{SW}}$ (dashed line) and $P_{\mathrm{RW1}}/P_{\mathrm{SW}}$ (dotted line) for a given gate error quantify the advantage conferred by the SW scheme. (c) The detuning down from the $P_{1/2}$ manifold corresponding to the solutions for the SW (solid line) and RW2 (dashed line) configurations at each $\epsilon_\mathrm{SPS}$.}
\label{fig:LS_result}
\end{figure*}

The laser power requirement has a complicated dependence on the error target in general, due to the nontrivial scaling of scattering rates with $\Delta$, relative magnitudes of the different error components, and their interplay with the gate drive strength. The full results of our numerical optimization are presented in the section below. Some insight can however be gained into the advantage offered by the SW scheme by calculating the gate error in the regime $|\Delta|\ll\omega_{\mathrm{fine}}$, where Raman scattering is dominated by scattering back to the ground state manifold. We first re-express $\epsilon_{\mathrm{SPS}}$ to emphasize its dependence on the beam powers, gate time and operating laser frequency. Neglecting the error due to motional decoherence, the gate error in schemes RW2 and RW1 can be written in the form \cite{ozeri2007errors}
\begin{equation}
\epsilon_{\mathrm{SPS}} = \mathcal{C}_{\mathrm{SPS}}\tau_{\mathrm{g}}\left(E^2_1+E^2_2\right)\left(\frac{\omega_{\mathrm{fine}}}{\Delta(\Delta+\omega_{\mathrm{fine}})}\right)^2, \label{eq:err-sps-simplified-RW}
\end{equation}
where all the dependence on the beam polarizations and the dipole coupling strengths is bundled into $\mathcal{C}_{\mathrm{SPS}}$. Similarly, the gate drive strengths in both LS and MS gates can be expressed as \cite{wineland2003quantum},
\begin{equation}
\Omega_{\mathrm{gate-drive}} = \mathcal{C}_{\mathrm{gate-drive}}E_1E_2\left|\frac{\omega_{\mathrm{fine}}}{\Delta(\Delta+\omega_{\mathrm{fine}})}\right|. \label{eq:Omega-gate-drive-simplified}
\end{equation}
The constraint \eqref{eq:Omega-gate-drive-constraint} on the gate drive strength together with \eqref{eq:err-sps-simplified-RW}, \eqref{eq:Omega-gate-drive-simplified} can be used to eliminate $\Delta$ and express the gate error in terms of the Raman field amplitudes:
\begin{equation}
\epsilon_{\mathrm{SPS}}=\frac{\mathcal{C}_{\mathrm{SPS}}}{\mathcal{C}^2_{\mathrm{gate-drive}}}\frac{\pi^2}{\tau_{\mathrm
{g}}}\frac{1}{\eta^2}\left(\frac{1}{E^2_1}+\frac{1}{E^2_2}\right) \label{eq:err-sps-simplified-RW-E}
\end{equation}
For a Gaussian beam with waists $w_{\mathrm{a}}$, $w_{\mathrm{b}}$, we have \cite{siegman1986lasers}, 
\begin{equation}
E^2_k=\frac{4}{\pi\epsilon_0c}\frac{P_k}{w_{\mathrm{a}}w_{\mathrm{b}}}.
\end{equation}
Since equal distribution of powers is optimal for the running-wave schemes, the minimum achievable error is,
\begin{align}
\epsilon_{\mathrm{SPS}} &= \frac{\pi^3\epsilon_0c}{4}\frac{\mathcal{C}_{\mathrm{SPS}}}{\mathcal{C}^2_{\mathrm{gate-drive}}}\frac{1}{\tau_{\mathrm
{g}}}\frac{w_{\mathrm{a}}w_{\mathrm{b}}}{\eta^2}\frac{4}{P_{\mathrm{tot}}}. \nonumber\\
&\qquad\qquad\qquad\quad \text{(RW1, RW2 Schemes)}\label{eq:err-sps-simplified-RW-P}
\end{align}
For the SW scheme, we can instead write,
\begin{equation}
\epsilon_{\mathrm{SPS}} = \mathcal{C}_{\mathrm{SPS}}\tau_{\mathrm{g}}\left(\eta^2\bar\alpha^2E^2_1+E^2_2\right)\left(\frac{\omega_{\mathrm{fine}}}{\Delta(\Delta+\omega_{\mathrm{fine}})}\right)^2, \label{eq:err-sps-simplified-SW}
\end{equation}
with the same gate drive strength as shown in eq.~\eqref{eq:Omega-gate-drive-simplified}. Here,
\begin{equation}
\bar\alpha^2\equiv\frac{\omega_{y_s}}{\omega_{y_c}}+\frac{3}{2}
\end{equation}
is close to the mean squared ion displacement averaged over the duration of the gate, but additionally accounts for contributions to the scattering rate from the undriven COM mode (see Appendix~\ref{App:scatter}). For the frequencies assumed here, $\bar\alpha\approx 1.6$. Then, we have,
\begin{align}
\epsilon_{\mathrm{SPS}}&=\frac{\pi^3\epsilon_0c}{4}\frac{\mathcal{C}_{\mathrm{SPS}}}{\mathcal{C}^2_{\mathrm{gate-drive}}}\frac{1}{\tau_{\mathrm
{g}}}\frac{1}{\eta^2}\left(\frac{1}{P_1}+\frac{\eta^2\bar\alpha^2}{P_2}\right)\nonumber\\
&\leq\frac{\pi^3\epsilon_0c}{4}\frac{\mathcal{C}_{\mathrm{SPS}}}{\mathcal{C}^2_{\mathrm{gate-drive}}}\frac{1}{\tau_{\mathrm
{g}}}\frac{w_{\mathrm{a}}w_{\mathrm{b}}}{\eta^2}\frac{(1+\eta\bar\alpha)^2}{P_{\mathrm{tot}}}, \nonumber\\
&\qquad\qquad\qquad\qquad\qquad\quad\text{(SW Scheme)}
\end{align}
with the optimal power fraction
\begin{equation}
p_1 = \frac{1}{1+\eta^{\mathrm{SW}}\bar\alpha}. \label{eq:optimal-power-dist}
\end{equation}
We quantify the advantage offered by the SW scheme by fixing the target gate error $\epsilon_{\mathrm{SPS}}$ and comparing the power required in the SW configuration to that in the conventional configurations. With the Lamb-Dicke parameters defined in eqs.~\eqref{eq:eta_RW1}-\eqref{eq:eta_SW} in the three configurations, we find,
\begin{align}
\frac{P_{\mathrm{tot}\text{,RW1}}}{P_{\mathrm{tot}\text{,SW}}}&=\frac{w_{\mathrm{a}}^{\mathrm{RW1}}w_{\mathrm{b}}^{\mathrm{RW1}}}{w_{\mathrm{a}}^{\mathrm{SW}}w_{\mathrm{b}}^{\mathrm{SW}}}\frac{4}{(1+\eta^{\mathrm{SW}}\bar\alpha)^2}, \label{eq:P-adv-RW1} \\
\frac{P_{\mathrm{tot}\text{,RW2}}}{P_{\mathrm{tot}\text{,SW}}}&=\frac{w_{\mathrm{a}}^{\mathrm{RW2}}w_{\mathrm{b}}^{\mathrm{RW2}}}{w_{\mathrm{a}}^{\mathrm{SW}}w_{\mathrm{b}}^{\mathrm{SW}}}\frac{8}{(1+\eta^{\mathrm{SW}}\bar\alpha)^2}. \label{eq:P-adv-RW2}
\end{align}
In our calculations we choose identical waists for the RW2 and SW schemes to enable direct comparison assuming the same focusing capabilities, so the prefactor in eq.~\eqref{eq:P-adv-RW2} is unity. For our choice of beam waists for the RW1 scheme the prefactor in eq.~\eqref{eq:P-adv-RW1} is 3.

\begin{figure*}[t]
\centering
\includegraphics[width=\linewidth]{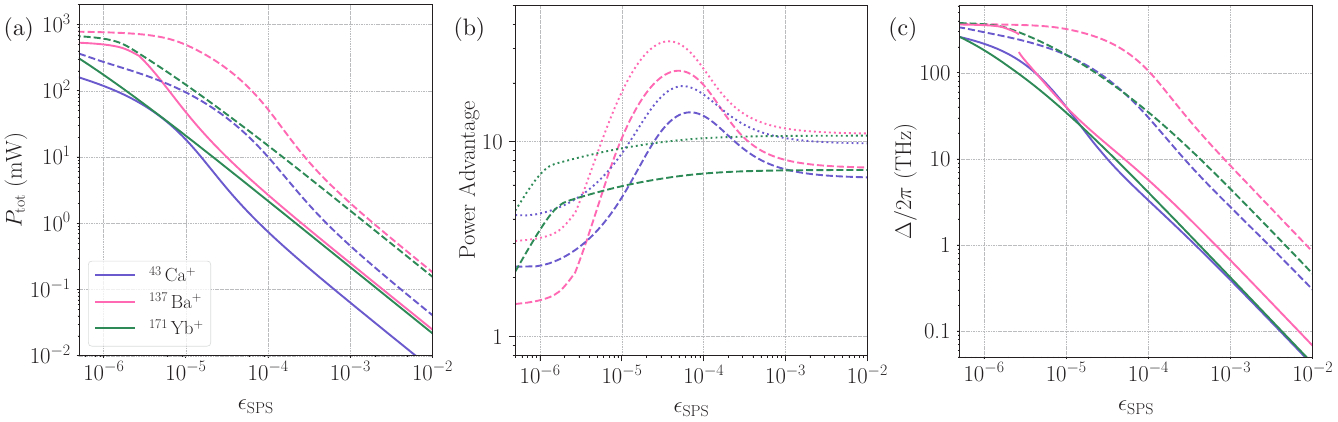}
\caption{Comparison of laser power requirement for MS gates with gate time $\tau_{\mathrm{g}}=50$ $\mu$s.  (a) Total laser power required given a target SPS-induced gate error $\epsilon_\mathrm{SPS}$, for the SW (solid line) and RW2 (dashed line) configurations. (b) The ratios of required total laser power $P_{\mathrm{RW1}}/P_{\mathrm{SW}}$ (dashed line) and $P_{\mathrm{RW2}}/P_{\mathrm{SW}}$ (dotted line) for a given gate error quantify the advantage conferred by the standing wave configuration. (c) The calculated detuning down from the $P_{1/2}$ manifold $\Delta$ is shown for the SW (solid line) and RW2 (dashed line) configurations.}
\label{fig:MS_result}
\end{figure*}

\begin{figure}[t]
\centering
\includegraphics[width=0.75\linewidth]{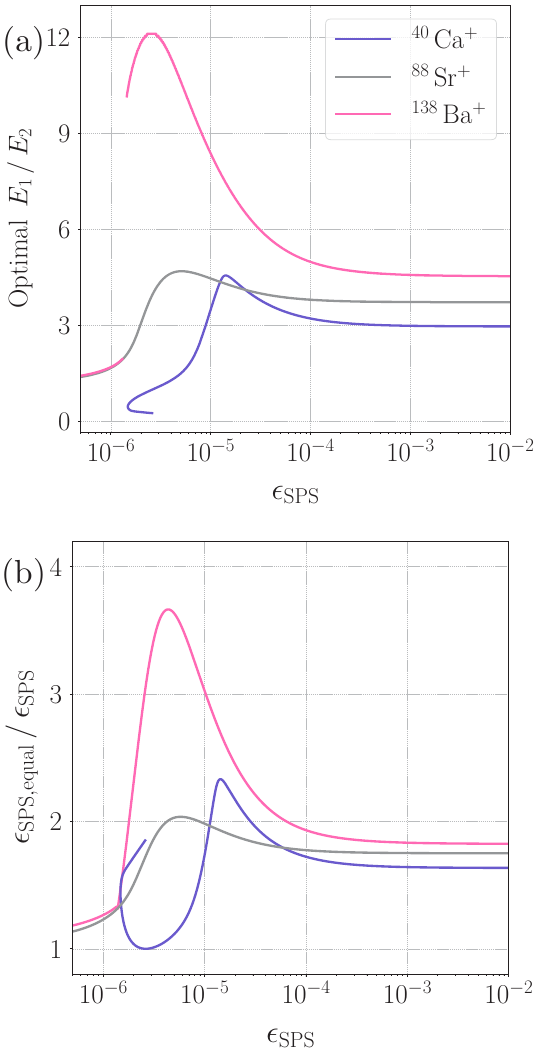}
\caption{Distribution of laser power between the Raman fields in the SW scheme for the LS gate. (a) The optimal ratio of field amplitudes of the SW and the RW required to achieve a target gate error $\epsilon_{\mathrm{SPS}}$. (b) To quantify the sensitivity to the power distribution, we calculate the ratio of the gate error for equal amplitudes in the SW and RW fields and the minimum achievable gate error, for fixed total power. The detuning in each case is set by constraint \eqref{eq:Omega-gate-drive-constraint}.} 
\label{fig:LS_sensitivity}
\end{figure}

The relatively weak dependence on $\eta$ in these expressions suggests the enhancements are fairly insensitive to ion species. These expressions largely account for the enhancements at low Raman detunings $\Delta$ observed in our full numerical optimization, which we turn to now. 

\subsection{Results}
We now compare the trade-off between the gate error and the laser power requirement in the three beam configurations considered here, for general detunings. For concreteness, we assume radial frequencies $\omega_{y_c}=2\pi\times 5.5$ MHz and $\omega_{y_s}=2\pi\times 5$ MHz and consider $\tau_{\mathrm{g}}=50$ $\mu$s for all species and configurations. The wavevectors of the gate beams in the RW2 and SW configurations are symmetric with respect to the ion positions, permitting the use of elliptical beam spots to maximize the intensity delivered for a given power in the beam; we consider using a narrow focus transverse to the trap axis, and a larger spot size along the trap axis to illuminate both ions. In our calculations, we assume that the beams for the RW2 and SW schemes have identical Gaussian field profiles with waist $w_{\mathrm{a}}=6$ $\mu$m along the direction with projection on the trap axis ($\mathbf{u_x}$), and $w_{\mathrm{b}}=2$ $\mu$m along $\mathbf{u_y}$, each measured perpendicular to the beam wavevector. In the RW1 configuration, we assume circular spots with waists $w_{\mathrm{a}}=w_{\mathrm{b}}=6$ $\mu$m.  

We present our results for a range of operating laser frequencies for the LS gate acting on ${}^{40}\mathrm{Ca}^+$, ${}^{88}\mathrm{Sr}^+$ and ${}^{138}\mathrm{Ba}^+$ in Fig.~\ref{fig:LS_result}, focusing here on laser frequencies that are red-detuned from the $P_{1/2}$ manifold (see Ref. \cite{moore2023photon} for a comparison between SPS rates for red detunings with respect to $P_{1/2}$, and blue detunings with respect to $P_{3/2}$). We see that for infidelities at the high end of the range considered here, the gate can be performed with about an order of magnitude less power in the SW configuration compared to the more conventional RW1 and RW2 configurations, with a significantly larger advantage in power requirement at lower errors. Among the species considered here, the advantage is most pronounced for ${}^{40}\mathrm{Ca}^+$. As we approach large detunings $\Delta$ to achieve smaller errors, the gate drive strength starts falling off at intermediate values $\Delta\approx\omega_{\mathrm{fine}}$. In the case of ${}^{40}\mathrm{Ca}^+$, due to its small fine-structure splitting, the drive strength starts falling off at detunings of a few ten THz while the scattering rate remains high. This forces us into the regime of large detunings of a few hundred THz, where the contribution of the qubit decoherence rate to $\epsilon_{\mathrm{SPS}}$ becomes negligible. We find that $\epsilon_{\mathrm{SPS}}$ in ${}^{40}\mathrm{Ca}^+$ is ultimately limited by the recoil error, but the large laser power requirement in this regime would be prohibitively difficult to access in practice. The inherently lower scattering error rate in the SW scheme effectively move the power-versus-error curve (Fig.~\ref{fig:LS_result}a) and the detuning-versus-error curve (Fig.~\ref{fig:LS_result}b) leftwards relative to the respective curves in the RW configurations for all three species considered. 

Our main results for the MS gate are shown in Fig.~\ref{fig:MS_result} for ${}^{43}\mathrm{Ca}^+$, ${}^{137}\mathrm{Ba}^+$ and ${}^{171}\mathrm{Yb}^+$; results for the power requirements in ${}^{9}\mathrm{Be}^+$, ${}^{25}\mathrm{Mg}^+$ and ${}^{87}\mathrm{Sr}^+$ are shown in Appendix~\ref{App:scatter}. The results are qualitatively similar to those for the LS gate. As we show in the Appendix, in the case of the lightest ion species, the predicted enhancements tend to be more modest due to the increased influence of recoil-induced error not suppressed by the SW  scheme, as clear from the trends for $\epsilon_\mathrm{qub}$ and $\epsilon_\mathrm{mot}$. The dramatic enhancement in power advantage for ${}^{40}\mathrm{Ca}^+$ is absent for the MS gate in ${}^{43}\mathrm{Ca}^+$; this can be attributed to a more favorable trade-off between MS gate drive strength and decoherence rate per unit power at intermediate detunings.  

The optimal power distribution in the SW configuration is highly asymmetrical; in the low-detuning limit \eqref{eq:optimal-power-dist}, around $90\%$ of the total power has to be sent to the SW to achieve minimum error. We plot the optimal distribution as a ratio of field amplitudes $E_1/E_2$ for a given target error $\epsilon_{\mathrm{SPS}}$ in Fig.~\ref{fig:LS_sensitivity}a. We find that the asymmetry in the optimal power distribution increases monotonically as we approach lower gate errors, before a nearly symmetric power distribution becomes optimal in the low-$\epsilon_{\mathrm{SPS}}$ regime. To get some insight into the sensitivity of the SW scheme to the asymmetric distribution of power, in Fig.~\ref{fig:LS_sensitivity}b we compare the error $\epsilon_{\mathrm{SPS, equal}}$ that would be accrued in the case of a symmetric power distribution, $p_1=1/2$, to the minimum achievable error $\epsilon_{\mathrm{SPS}}$. Comparing this to Fig.~\ref{fig:LS_result}b, we find that the significant power advantage conferred by the SW configuration is robust to the power distribution between the SW and the RW for $\epsilon_{\mathrm{SPS}}$ down to order $10^{-6}$ for the gate parameters considered here. 

We bound the possible inaccuracy resulting from our treatment of Rayleigh scattering in terms of separate impacts on internal and motional state despite their entanglement during the gate, by considering a ``pessimistic" limit  in which we assume every Rayleigh scattering event induces a gate error. As shown in Appendix~\ref{App:scatter}, even in this extreme limit the predicted enhancements are minimally affected except for gates driven at very large detunings and powers and at the lowest infidelities. 

\section{Discussion/Conclusion}
Our work extends previous treatments of SPS errors in trapped-ion quantum logic to account for simple instances of structured light driving fields. In the SW scheme considered here SPS arises only due to sideband absorption from the SW beam, and becomes sensitive to the motional excursion during the gate. Our calculations indicate that for optimal power distributions between the SW and RW fields, required drive power is reduced by approximately an order of magnitude for low detunings resulting in gate errors of $10^{-3} - 10^{-2}$, with further enhancement at higher fidelities in the $10^{-4}-10^{-5}$ regime depending on ion species and gate scheme. These enhancements are predicted across a range of ion species considered for both LS and MS gates acting on Zeeman and approximate low-field clock states. The details of all the ion species considered here are presented in Table~\ref{tab:my-table}. In the low Raman detuning regime, the weak dependence of the power enhancement factor on the Lamb-Dicke parameter results in comparable advantage across a range of ion species. However, as calculated here, the fact that the SW field brings no significant advantage in motional decoherence due to recoil effects means that ions with low ion mass suffer from a relatively larger recoil-associated error in our scheme.

\begin{table*}[btp]
\centering
\setlength{\tabcolsep}{5pt}
\begin{tabular}{@{}lllllll@{}}
\toprule
& $\mathrm{Be}$& $\mathrm{Mg}$& $\mathrm{Ca}$& $\mathrm{Sr}$& $\mathrm{Ba}$& $\mathrm{Yb}$\\\midrule
LS isotope&&& $^{40}\mathrm{Ca}^+$&$^{88}\mathrm{Sr}^+$&$^{138}\mathrm{Ba}^+$&\\
MS isotope, Nuclear spin& ${}^{9}\mathrm{Be}^+$, $I=3/2$ & $^{25}\mathrm{Mg}^+$, $I=5/2$ & $^{43}\mathrm{Ca}^+$, $I=7/2$ & $^{87}\mathrm{Sr}^+$, $I=9/2$ & $^{137}\mathrm{Ba}^+$, $I=3/2$ & $^{171}\mathrm{Yb}^+$, $I=1/2$ \\
$\lambda_{S_{1/2}\leftrightarrow P_{1/2}}$ (nearest nm) & 313& 280 & 397& 421& 493& 369\\
$\gamma_{P_{3/2}}/2\pi$ (MHz) & 18.0& 41.4 & 21.6& 22.4&17.7& 25.8\\
$\alpha_{D_{3/2}}$ & & & 0.0063& 0.0066&0.0379& 0.0017\\
$\alpha_{D_{5/2}}$ & & & 0.0535& 0.0577&0.2604& 0.0108\\
$\omega_{\mathrm{fine}}/2\pi$ (THz)&0.2&2.7&6.7&24.0&50.7&99.9 \\\bottomrule
\end{tabular}
\caption{Atomic species and parameters used in calculations. $\gamma_{P_{3/2}}$ is the natural linewidth of the $P_{3/2}$ manifold and $\alpha_{D_{3/2}},\alpha_{D_{5/2}}$ are the branching rates of scattering from $P_{3/2}$ to the $D$ levels. Data taken from \cite{NIST_ASD_2025}.}
\label{tab:my-table}
\end{table*}

To take a representative example of the advantage predicted, an MS gate with $\tau_g = 50$ $\mu$s and a total $\epsilon_\mathrm{SPS} = 10^{-4}$ using $^{137}$Ba$^+$ requires approximately 50-70 mW of total optical power in the RW schemes for the beam profiles and gate parameters assumed in our work, which with the SW drive scheme is reduced to approximately 3 mW. This power enhancement would allow for considerably larger numbers of parallel gate zones implemented in a multiplexed trap with acceptable total optical powers delivered, or alternatively, significantly faster gates for a given total optical power. For LS gates with $^{40}$Ca$^+$, the results in Fig.~\ref{fig:LS_result}a show that in the SW scheme gates with $<10^{-4}$ error are achievable with sub-mW level optical powers; in the comparable RW schemes, the power requirements quickly become prohibitive at these infidelities.  

With respect to gate speed, we note that in addition to the reduced $\epsilon_\mathrm{SPS}$ allowing faster operation for a given target infidelity and available optical power, the SW drive schemes suppress coherent driving due to the ``carrier" term in the interaction Hamiltonian (see Appendix A) \cite{mehta2019towards, saner2023breaking}. This suppresses the non-commuting carrier term that impedes high-fidelity MS gate operation for $\tau_g$ approaching the motional mode period, and allows for minimized pulse shaping to account for off-resonant couplings in LS gates \cite{steane2014pulsed}. The further suppression of even-order expansion terms also suppresses out-of-LD effects that are problematic to compensate for in very fast gates \cite{schafer2018fast}. The simultaneous power advantage together with suppressed off-resonant coherent couplings in the proposed scheme hence appears promising for practical implementation of fast laser-based gates. The breakdown of the validity of the LD approximation in this regime would require modifications to the SW scattering error calculations -- presented here within the LD regime -- which would be an interesting extension for future work.

While our calculations necessarily focus on a subset of possible qubit species and encodings, similar enhancements are expected to hold for other encodings, including finite-field clock states as well as metastable \cite{allcock2021omg, moore2023photon, wang2025experimental} and optical qubits \cite{sawyer2021wavelength} which may operate in vastly different wavelength regimes. We have focused on red detunings from the $P_{1/2}$ states; for blue detunings from $P_{3/2}$, the scattering error has a lower bound \cite{moore2023photon} and tends to be less favorable, although our expressions can straightforwardly be used to predict performance for these detunings.

The primary challenge we anticipate with the SW scheme is associated with the motional frequency shift that results from the spatial curvature in the optical intensity (and associated AC Stark shift) experienced by the ion. For both the MS and LS schemes considered here, this motional frequency shift is however independent of the ion's internal state. It is furthermore static in time, avoiding potential squeezing dynamics that can introduce infidelites \cite{vasquez2024state}. It is hence an effect that can in principle be straightforwardly calibrated without introducing nontrivial additional dynamics. The magnitude of this shift is calculated in Appendix~\ref{App:gateham} for the gate parameters considered here, with the resulting shifts in radial frequencies plotted in Fig.~\ref{fig:freq_shift}. The shift, at the few kHz level for modest gate errors, is appreciable compared to the gate detuning (20 kHz in this work) and increases in the high-fidelity regime, and hence is critical to account for in high-fidelity implementations. 

The SW scheme also introduces more demanding positioning requirements than typical gate configurations. However recent experiments with ion control in phase-stable standing waves sourced from integrated optics \cite{vasquez2023control, clements2026sub} suggest that few nm-level positioning control is practical in such settings even with present understanding of materials for stray field shielding \cite{de2021materials, brown2021materials}. We show in Appendix \ref{App:scatter} that gate fidelities in the SW scheme are minimally affected by expected positioning inaccuracies. How challenging it is to realize such stable positioning with the required optical powers emitted through surface openings and the potential associated evolution of stray charge \cite{harlander2010trapped, wang2011laser} is a key question for experimental realization. We have assumed the same $\theta_z$ for all beams for simplicity here, but minor modifications can relax some constraints; for example we note that the additional positioning requirements for ions along the axial direction for the RW2 and SW schemes considered here may be relaxed by use of a vertically emitted $\mathbf{E_2}$ field with no wave-vector projection along the axis. 

Structured light profiles other than SW nodes could also be used to generate the gate interaction presented here. For example, first-order Hermite-Gauss modes have the same first-order field gradient along the nodal line and would permit SPS suppression at the intensity node in a similar fashion as in the SW scheme. An entangling gate implementation employing such a configuration to drive axial modes of motion in ion chains has been demonstrated recently \cite{faorlin2026entangling}. The effective single-ion LD parameter for such a configuration with a transverse gradient along the $y$-direction, $\eta_\mathrm{HG} = \frac{2}{w_y} y^{(0)}$ \cite{west2021tunable} is however challenging to increase to that achieved for a SW $\eta_\mathrm{SW} = \frac{2\pi}{\lambda} y_0$ except for subwavelength waists $w_y$; hence for practical focuses it appears challenging to achieve a gain in power requirement with such a configuration. 

The concepts and analysis presented here indicate that the use of more sophisticated structured light profiles \cite{schmiegelow2012light, verde2023trapped} in Raman interactions offer interesting opportunities to mitigate SPS beyond the simple case considered here. For instance, SW-like field profiles offer a means to generating scatter-suppressed tweezer interactions and squeezing terms in the Hamiltonian that can furthermore be made state-dependent \cite{katz2022n,schwerdt2025optical,mazzanti2021trapped, burd2024experimental}. Carrier-nulling by positioning the ions at intensity nulls suppresses the large AC Stark shifts that can limit the performance of RW-based schemes utilizing such interactions \cite{schwerdt2025optical}, which may be particularly advantageous in squeezing and higher-order interactions in view of the larger drive strengths typically required. Our analysis of SPS here also informs studies of stimulated-Raman interactions coupling to atomic motion in structured light fields more generally. For instance, Raman sideband cooling is often utilized to perform ground-state cooling in both trapped-ion and neutral atom QIP systems \cite{monroe1995resolved, thompson2013coherence,  kaufman2012cooling, moses2023race, jenkins2022ytterbium}. The analysis for carrier-suppressed stimulated Raman interactions may assist in improving upon the cooling limit and speed in such systems \cite{reimann2014carrier}.

Our work presents a simple, experimentally realizable configuration for entangling gates acting on a range of ground-state qubit encodings, that leverages the phase-stability afforded by scalable approaches to optical control to suppress spontaneous photon scattering errors and the associated limits on achievable laser gate fidelities. The calculations presented here suggest significant potential for structured light profiles in stimulated Raman optical interactions more generally. In offering routes to high-fidelity laser gates at lower powers and/or higher speeds, we expect the analysis presented here will significantly impact future architectures for laser-based trapped-ion control particularly in large-scale systems. 

\textit{Note added in proof: Recent works \cite{cui2025transverse, mai2025scalable} demonstrate the use of transverse field gradients to drive entangling gates using axial motional modes in ion chains.}

\section*{Acknowledgments}
We thank Wes Campbell, Hartmut Haeffner, Jonathan Home, Eric Hudson, Thomas Monz, Brian Sawyer, and Jeff Thompson for helpful discussions. We acknowledge support from an NSF CAREER Award (No. 2338897); IARPA and the Army Research Office, under the Entangled Logical Qubits program, Cooperative Agreement Number W911NF-23-2-0216; the NSF NQVL program (No. 2435382); the Alfred P. Sloan Foundation; and Cornell University. The views and conclusions contained in this document are those of the authors and should not be interpreted as representing the official policies, either expressed or implied, of IARPA, the Army Research Office, or the U.S. Government. The U.S. Government is authorized to reproduce and distribute reprints for Government purposes notwithstanding any copyright notation herein.

\section*{Data Availability}
The data that support the findings of this article are openly available \cite{repo}.

\appendix

\section{Derivation of coherent drive} \label{App:gateham}

Here, we will derive the LS and MS gate interactions for implementations with ground state qubits. We will begin by reviewing the general form of a stimulated Raman interaction in a three-level ladder system and then show how the familiar forms of the LS and MS gate Hamiltonians arise for particular Raman field configurations. Throughout this section, we will refer to the beam configurations shown in Fig. \ref{fig:beam_geometries_schematic} in the main text. 

\subsection{Effective interaction in a ladder system}
Consider an ion where two sub-levels $\ket{\downarrow}, \ket{\uparrow}$ of the ground state manifold are coupled to an excited state $\ket{e}$ by dipole transitions. A pair of Raman fields, $\mathbf{E_1}$ and $\mathbf{E_2}$ drive transitions between the states. We express the Raman fields  in the form
\begin{equation}
\mathbf{E_i}(\mathbf{r},t)=\boldsymbol{\epsilon}_i(\mathcal{E}_i(\mathbf{r})e^{-i\omega_it}+c.c.),
\end{equation}
where $\mathcal{E}_i(\mathbf{r})=E_ie^{i\mathbf{k_i}\cdot\mathbf{r}+i\phi_i}/2$ in the case of a plane wave field and $\mathcal{E}_i(\mathbf{r})=E_i\sin(\mathbf{k_i}\cdot\mathbf{r}+\phi_i)/\sqrt
{2}$ in the case of a standing wave (SW) superposition. The factor of $\sqrt{2}$ in the latter normalizes the scalar amplitude $E_i$ so that the SW superposition uses the same total laser power as a plane-wave Raman field with amplitude $E_i$. 

\begin{figure*}[t]
\centering
\includegraphics[width=0.85\linewidth]{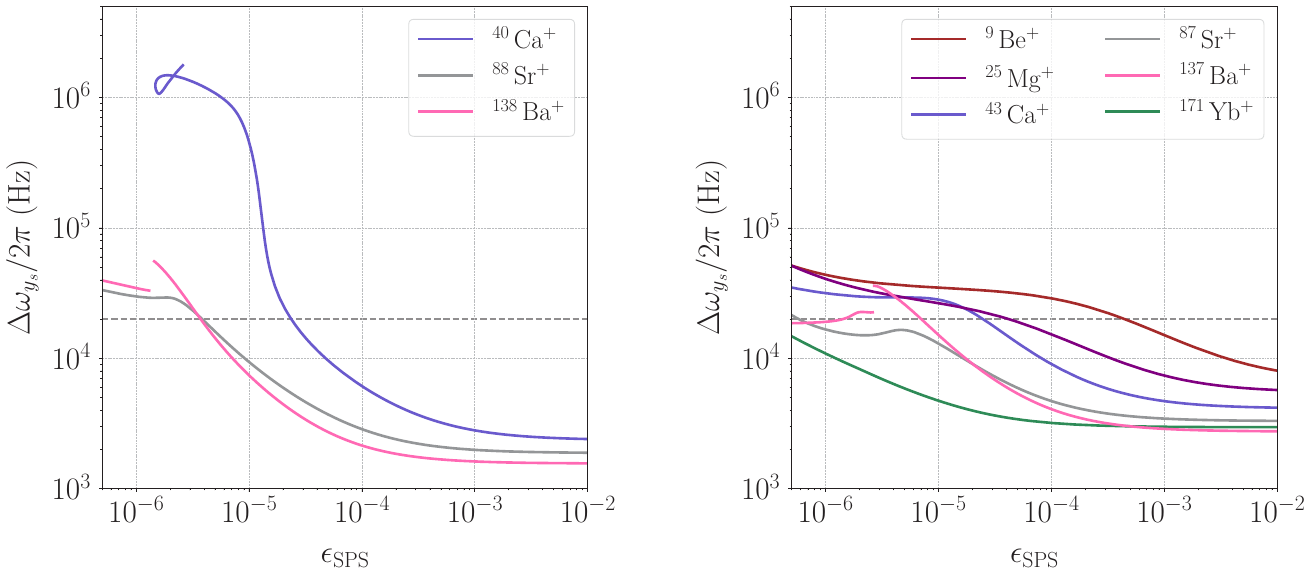}
\caption{The frequency shift $\Delta\omega_{y_s}$ for target gate error $\epsilon_{\mathrm{SPS}}$ from spontaneous photon scattering in the SW beam configuration, for LS gates (left) and MS gates (right). The dashed line marks the gate detuning $\delta/2\pi=20$ kHz for the $\tau_{\mathrm{g}}=50$ $\mu$s gates considered here.}
\label{fig:freq_shift}
\end{figure*}

The detunings of the Raman beams from the ${}^2P$ levels that we consider are much larger than the other frequency scales in the problem, namely (i) the natural linewidth of the excited states, (ii) the gate speed $\tau_{\mathrm{g}}^{-1}$, (iii) the dipole transition Rabi rates connecting the ground and excited states and (iv) the qubit splitting $\omega_0$. We therefore adiabatically eliminate the excited state $\ket{e}$ to get the effective interaction $\hat H_\mathrm{int}=\sum_{s=\downarrow,\uparrow}\sum_{s'=\downarrow,\uparrow}H_{ss'}\ket{s}\bra{s'}$ due to the stimulated Raman process. In the interaction picture with respect to the bare Hamiltonian of the internal and the motional degrees of freedom, we have \cite{wineland1998experimental},
\begin{widetext}
\begin{align}
H_{ss'}=\sum_{i=1}^2 \frac{1}{\hbar}|\mathcal{E}_i(\mathbf{r})|^2&\bra{s}\mathbf{d}\cdot\boldsymbol{\epsilon}_i\ket{e}\bra{e}\mathbf{d}\cdot\boldsymbol{\epsilon}_i\ket{s'}\left(\frac{1}{\omega_{eg}-\omega_{\mathrm{las}}}+\frac{1}{\omega_{eg}+\omega_{\mathrm{las}}}\right)e^{i\omega_{ss'}t}+ h.c.\nonumber\\
&+\frac{1}{\hbar}\mathcal{E}_1(\mathbf{r})\mathcal{E}_2^*(\mathbf{r})\left(\frac{\bra{s}\mathbf{d}\cdot\boldsymbol{\epsilon}_2\ket{e}\bra{e}\mathbf{d}\cdot\boldsymbol{\epsilon}_1\ket{s'}}{\omega_{eg}-\omega_{\mathrm{las}}}+\frac{\bra{s}\mathbf{d}\cdot\boldsymbol{\epsilon}_1\ket{e}\bra{e}\mathbf{d}\cdot\boldsymbol{\epsilon}_2\ket{s'}}{\omega_{eg}+\omega_{\mathrm{las}}}\right)e^{i(\omega_{ss'}+\omega_2-\omega_1)t}+h.c.\label{eq:H_ss'}
\end{align}    
\end{widetext}
Here, $\mathbf{\hat d}=e\mathbf{\hat r_{el}}$ is the dipole operator, $\omega_{\mathrm{las}}$ is the average optical frequency of the Raman fields and $\omega_{eg}$ is the splitting between the ground state and the excited states. We have invoked the rotating wave approximation (RWA) to drop terms rotating at optical frequencies. In deriving the LS and MS interactions, we will invoke the RWA again to drop terms rotating much faster than the gate speed $\tau_{\mathrm{g}}^{-1}$.

Each term in \eqref{eq:H_ss'} may be thought of as being indexed by $(i,i')$, arising from stimulated absorption from beam $i$ and stimulated emission from beam $i'$. Terms of the form $(i,i)$ produce an AC Stark shift on the qubit states. We use linearly-polarized fields throughout, so the shift is the same on both levels in the Zeeman qubit \cite{wineland2003quantum}. Moreover, the AC Stark shift is always equal on the clock-qubit states \cite{lee2005phase}. In the running wave configurations RW2 and RW1, we absorb the Stark shift into our definitions of the energy of the qubit levels. In the SW configuration, the curvature of the resulting SW optical potential shifts the frequency of the motional modes. We absorb this shift into the gate detuning $\delta$ (see discussion below). 

The terms $(i,i')$ in \eqref{eq:H_ss'} with $i\neq i'$ drive the gate either by generating an additional, time-varying, Stark shift (LS gates) or by coupling the two qubit levels (MS gates). The strength of the gate interaction is proportional to the two-photon Rabi frequency,
\begin{align}
\Omega_{s,s'}=g_1g_2\sum_k\Bigg(&\frac{\bra{s}\hat{\mathbf{r}}\mathbf{_{el}}\cdot\boldsymbol{\epsilon}_2\ket{k}\bra{k}\hat{\mathbf{r}}\mathbf{_{el}}\cdot\boldsymbol{\epsilon}_1\ket{s'}}{\mu^2(\omega_{kg}-\omega_{\mathrm{las}})} \nonumber\\
&\qquad+\frac{\bra{s}\hat{\mathbf{r}}\mathbf{_{el}}\cdot\boldsymbol{\epsilon}_1\ket{k}\bra{k}\hat{\mathbf{r}}\mathbf{_{el}}\cdot\boldsymbol{\epsilon}_2\ket{s'}}{\mu^2(\omega_{kg}+\omega_{\mathrm{las}})}\Bigg)
\end{align}
where we now sum the contributions from multiple excited states; here, $k$ indexes all sublevels in the $P_{1/2}$ and $P_{3/2}$ manifolds. $\omega_{kg}$ is the mean angular frequency of the transition between the ground state and the excited state, $g_i=eE_i\mu/2\hbar$ and $\mu$ is the largest dipole matrix element connecting the qubit levels to the manifold of excited states. For species with hyperfine structure for instance,
\begin{align*}
\mu=&|\bra{F'=I+3/2, m'_F=I+3/2} \\
&\qquad\qquad\mathbf{\hat r_{el}\cdot \mathbf{u_{\sigma_+}}}\ket{F=I+1/2,m_F=I+1/2}|.
\end{align*}

\begin{figure*}[t]
\centering
\includegraphics[width=\linewidth]{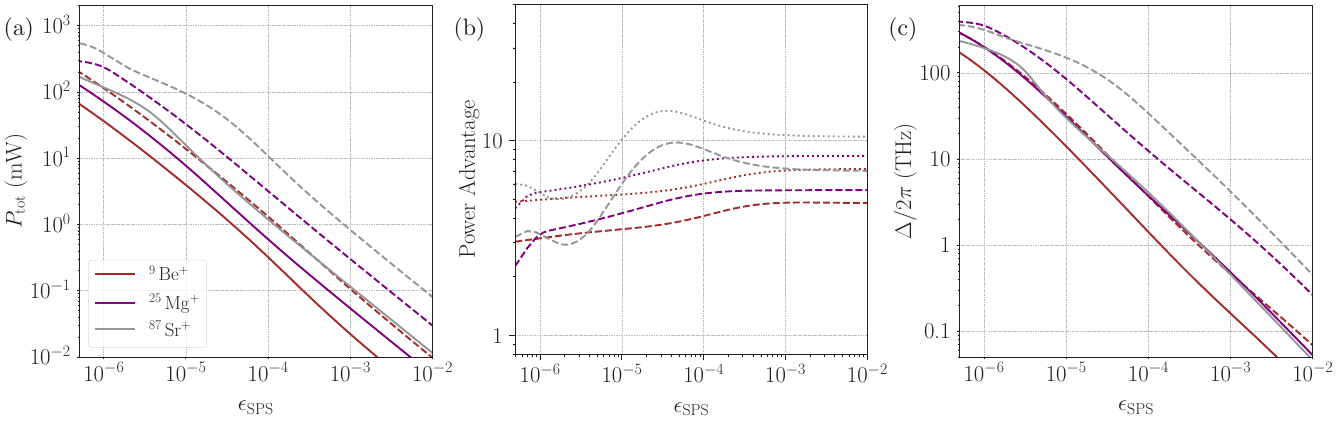}
\caption{Comparison of laser power requirement for MS gates with gate time $\tau_{\mathrm{g}}=50$ $\mu$s for species not presented in Fig.~\ref{fig:MS_result}. (a) Total laser power required given a target SPS-induced gate error $\epsilon_\mathrm{SPS}$, for the SW (solid line) and RW2 (dashed line) configurations. (b) The ratios of required total laser power $P_{\mathrm{RW1}}/P_{\mathrm{SW}}$ (dashed line) and $P_{\mathrm{RW2}}/P_{\mathrm{SW}}$ (dotted line) for a given gate error quantify the advantage conferred by the standing wave configuration. (c) The calculated detuning down from the $P_{1/2}$ manifold $\Delta$ is shown for the SW (solid line) and RW2 (dashed line) configurations.}
\label{fig:App_MS_result}
\end{figure*}

\subsection{LS and MS gate Hamiltonians}
For LS gates on Zeeman qubits, we set the Raman difference frequency $\omega_2-\omega_1=\omega_{y_s}+\delta$ with $|\delta|\ll\omega_{y_s}$ to selectively drive the stretch mode in the $y$-direction, $y_s$, in a near-resonant fashion. Then the terms $s\neq s'$ in \eqref{eq:H_ss'} effecting qubit-flips can all be dropped by the RWA. Working within the Lamb-Dicke regime, we expand the interaction to first order in the ion position operator $\hat{\mathbf{r}}\mathbf{_j}$, which we express as,
\begin{equation}
\hat{\mathbf{r}}\mathbf{_j}=\mathbf{r_j^{(eq)}}+\hat{\mathbf{q}}\mathbf{_j}, \label{eq:ion_position_operator}
\end{equation}
with $\hat{\mathbf{q}}\mathbf{_j}$ denoting the displacement operator of ion $j$ with equilibrium position $\mathbf{r_j^{(eq)}}$. We then obtain the gate interaction by summing the effective interaction \eqref{eq:H_ss'} over the two ions. 

In the RW1 and RW2 configurations, the Raman fields are plane waves with spatial profiles $\mathcal{E}_i=E_ie^{i\mathbf{k_1}\cdot\mathbf{r}}/2$. The coupling to motion in this case is proportional to the difference field gradient of the Raman beams. The LS gate interaction takes the following form:
\begin{align}
\frac{\hat H_{\mathrm{int}}^{\mathrm{LS}}}{\hbar} &= \sum_j\sum_{s_j}\Omega_{s_js_j}\left(1+i\mathbf{\Delta k}\cdot\hat{\mathbf{q}}\mathbf{_j}\right)\times\nonumber\\
&\qquad\qquad\qquad e^{i(\omega_{y_s}+\delta)t+i\phi_{\mathrm{m}}}\ket{s_j}\bra{s_j}+h.c. \label{eq:H-LS-without-RWA}
\end{align}
where $\mathbf{\Delta k}=\mathbf{k_1}-\mathbf{k_2}$ is the difference wavevector and $\phi_{\mathrm{m}}=\phi_1-\phi_2+\mathbf{\Delta k}\cdot \mathbf{r_j^{(eq)}}$ is the phase of the force. In configuration RW1, $\mathbf{\Delta k}$ lies along the $y$-axis, so the phase of the force is automatically the same on the two ions assuming that they lie along the trap axis. In configuration RW2, $\mathbf{\Delta k}$ has a projection along the trap axis, so the ion spacing along this axis has to be chosen to be $\mathbf{\Delta k}\cdot\mathbf{u_x}({x^{\mathrm{(eq)}}_{1}}-x^{\mathrm{(eq)}}_{2})=2\pi n$ for some integer $n$ in order to drive the gate most efficiently \cite{ballance2016high}.  

We now expand the displacement operators $\hat{\mathbf{q}}\mathbf{{}_j}$ in the eigenbasis of the shared motional modes of the two ions \cite{James1998Quantum}. In the lab frame, we have along the $y$-direction,
\begin{equation}
\hat{q}_{y,j}=\frac{1}{\sqrt{2}}b_{y_c,j}y_{c}^{(0)}(\hat a_{y_c}+\hat a_{y_c}^\dagger)+\frac{1}{\sqrt{2}}b_{y_s,j}y_{s}^{(0)}(
\hat a_{y_s}+\hat a_{y_s}^\dagger).
\end{equation}
where the expansion coefficients $b_{y_c,j}=1$ and $b_{y_s,j}=(-1)^j$ and $y^{(0)}_l\equiv \sqrt{\hbar/2m\omega_{l}}$ is the RMS extent of the motional ground state in mode $l$. We also define the operators
\begin{equation}
\hat y_l \equiv y^{(0)}_l(\hat a_l+\hat a^\dagger_l).
\end{equation}
In the interaction picture with respect to the bare Hamiltonian, the annihilation (creation) operators are time-dependent; $\hat a_{l}(t)=\hat a_l e^{-i\omega_l t}$ ($\hat a^\dagger_{l}(t)=\hat a^\dagger_l e^{i\omega_l t}$).

Neglecting all fast-rotating terms in \eqref{eq:H-LS-without-RWA}, we arrive at the gate Hamiltonian,
\begin{align}
\frac{\hat H_{\mathrm{int}}^{\mathrm{LS}}}{i\hbar}&=\sum_{s_1,s_2}\eta\big(\sum_j b_{y_s,j}\Omega_{s_js_j}\big)\,\hat a e^{i\delta t+i\phi_{\mathrm{m}}}\ket{s_1s_2}\bra{s_1s_2}+h.c.\nonumber\\
&=\sum_{s_1,s_2}\eta\left(\Omega_{s_1s_1}-\Omega_{s_2s_2}\right)\,\hat a e^{i\delta t+i\phi_{\mathrm{m}}}\ket{s_1s_2}\bra{s_1s_2}+h.c.\nonumber \\ 
&=2\eta\Omega_{\downarrow\downarrow}\frac{\hat\sigma_{z,1}-\hat\sigma_{z,2}}{2}\,\hat a e^{i\delta t+i\phi_{\mathrm{m}}}+h.c.\label{eq:H-LS-RWA}
\end{align}
where in going to the last line, we have used the fact that $\Omega_{\uparrow\uparrow}=-\Omega_{\downarrow\downarrow}$ for Zeeman qubits driven by crossed linear polarizations. Here we define the Lamb-Dicke parameter $\eta = \mathbf{\Delta k}\cdot\mathbf{u_y}y_{\mathrm{s}}^{(0)}/\sqrt{2}$ and denote the annihilation (creation) operators of the mode $y_s$ by $\hat a$ ($\hat a^{\dagger}$). For the two running wave beam configurations, we have, $\eta^{\mathrm{RW1}}=k_{\mathrm{las}}y_{s}^{(0)}\sin\theta_z$ and $\eta^{\mathrm{RW2}}=\eta^{\mathrm{RW1}}/\sqrt{2}$ (our results assume $\theta_z=60\degree$). 

\begin{figure*}[t]
\centering
\includegraphics[width=\linewidth]{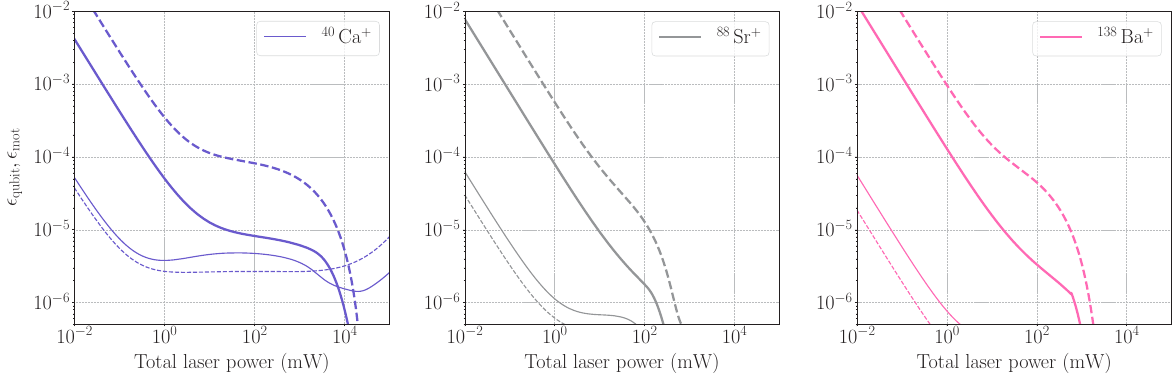}
\caption{Minimum achievable qubit decoherence and motional contributions to the LS gate error as a function of available total laser power. Solid (dashed) lines show calculated error in the SW (RW2) configuration. Thick (thin) lines denote the qubit decoherence (motional) contribution to the total gate error $\epsilon_{\mathrm{SPS}}$.}
\label{fig:LS_qubit_plus_recoil}
\end{figure*}

In the SW configuration, the Raman field $\mathbf{E_1}$ is a SW superposition with spatial profile $\mathcal{E}_1=E_1\sin(\mathbf{k_1}\cdot \mathbf{r})/\sqrt{2}$ and $\mathbf{E_2}$ is a running wave with spatial profile $\mathcal{E}_2=E_2e^{i\mathbf{k_2}\cdot \mathbf{r}}/2$, so the gate interaction, in analogy to \eqref{eq:H-LS-without-RWA}, is,
\begin{align}
\frac{\hat H_{\mathrm{int}}^{\mathrm{LS}}}{\hbar} &= \sum_j\sum_{s_j}\sqrt{2}\Omega_{s_js_j}\mathbf{k_{1}}\cdot\hat{\mathbf{q}}\mathbf{_j}\left(1-i\mathbf{k_2}\cdot\hat{\mathbf{q}}\mathbf{_j}\right)\times\nonumber\\
&\qquad\qquad\qquad e^{i(\omega_{y_s}+\delta)t+i\phi_{\mathrm{m}}}\ket{s_j}\bra{s_j}+h.c., \label{eq:H-LS-SW-without-RWA}
\end{align}
where we assume that the phase $\phi_1$ of the SW superposition is such that $y=0$ is a SW null. The phase of the force in this case is $\phi_{\mathrm{m}}=-\phi_2-\mathbf{k_2}\cdot\mathbf{r_j^{(eq)}}$, so the ions have to be spaced along the trap axis by $\mathbf{k_2}\cdot\mathbf{u_x}({x^{(eq)}_{1}}-x^{(eq)}_{2})=2\pi n$ for some $n$ to most efficiently excite the motion. Note that we for simplicity have assumed all beams at the same angle from normal $\theta_z$, but if $\mathbf{E_2}$ propagated along the vertical with zero $k_x$ component, $\phi_m$ would be insensitive to axial position. Note also that the lowest-order terms in eq.~\ref{eq:H-LS-SW-without-RWA} are first-order in $\mathbf{\hat q_j}$, in contrast to eq.~\ref{eq:H-LS-without-RWA}, reflecting the suppression of the coherent carrier drive term \cite{mehta2019towards, saner2023breaking}.

Upon expanding the ion displacements in the motional mode eigenbasis and dropping all fast-rotating terms, we arrive at the same expression as \eqref{eq:H-LS-RWA} up to an overall phase, and with $\eta^{\mathrm{SW}}=\eta^{\mathrm{RW1}}=k_{\mathrm{las}}y_{s}^{(0)}\sin\theta_z$. Here the Lamb-Dicke parameter is set not by $\mathbf{\Delta k}$ but by the effective wavevector of the SW. 

For MS gates acting on clock qubits, we choose the `phase-insensitive' arrangement \cite{halijan2005phase} as a particular case; we send a tone at $\omega_1=\omega_{\mathrm{las}}$ with phase $\phi_1$ in $\mathbf{E_1}$ and use a superposition of two tones, $\omega_{2,\mathrm{b}}=\omega_{\mathrm{las}}+\omega_0+\omega_{y_s}+\delta$ and $\omega_{2,\mathrm{r}}=\omega_{\mathrm{las}}-\omega_0+\omega_{y_s}+\delta$, with phases $\phi_{2,\mathrm{b}}$ and $\phi_{2,\mathrm{r}}$ respectively, in $\mathbf{E_2}$. It is most efficient to distribute power equally between the two tones. The gate interaction, then, is given by the pairwise sum of the contributions \eqref{eq:H_ss'} for the three tones, summed over the two ions. Assuming that the static AC Stark shifts have been absorbed into the definitions of the qubit energy levels and/or the gate detuning $\delta$, the gate Hamiltonian for the running wave beam configurations takes the form,
\begin{align}
\frac{\hat H_{\mathrm{int}}^{\mathrm{MS}}}{\hbar}=&\sum_j\Omega_{\downarrow\uparrow}\left(1+i\mathbf{\Delta k}\cdot\hat{\mathbf{q}}\mathbf{_j}\right)e^{i(\omega_{y_s}+\delta) t+i\phi^{(j)}}\hat{\sigma}_{-,j}\nonumber\\
&\,+\sum_j\Omega_{\downarrow\uparrow}\left(1-i\mathbf{\Delta k}\cdot\hat{\mathbf{q}}\mathbf{_j}\right)e^{-i(\omega_{y_s}+\delta) t-i\phi'{}^{(j)}}\hat{\sigma}_{-,j}\,+\,h.c.
\end{align}
where for ion $j$, $\phi^{(j)}\equiv\phi_{1}-\phi_{2,\mathrm{b}}+\mathbf{\Delta k}\cdot\mathbf{r_j^{(eq)}}$ with $\phi'{}^{(j)}$ defined similarly, replacing $\phi_{2,\mathrm{b}}$ by $\phi_{2,\mathrm{r}}$. Expanding the ion displacements in the eigenbasis of the motional modes as before and dropping all fast-rotating terms, we are left with
\begin{align}
\frac{\hat H_{\mathrm{int}}^{\mathrm{MS}}}{i\hbar}=&\sum_j\frac{1}{\sqrt{2}}\eta\Omega_{\downarrow\uparrow}(-1)^j\left(e^{i\phi'{}^{(j)}}\hat\sigma_{-,j}+e^{i\phi^{(j)}}\hat\sigma_{+,j}\right)\,\hat ae^{i\delta t} \nonumber\\
&\qquad\qquad\qquad\qquad\qquad\qquad\qquad\qquad\;\;+ h.c. \nonumber\\
=& \sum_j\frac{1}{\sqrt{2}}\eta\Omega_{\downarrow\uparrow}(-1)^j\left(e^{i\phi_s}\hat\sigma_{-,j}+e^{-i\phi_s}\hat\sigma_{+,j}\right)\,\hat a e^{i\delta t+i\phi^{(j)}_m} \nonumber \\
&\qquad\qquad\qquad\qquad\qquad\qquad\qquad\qquad\;\;+h.c. \nonumber \\
=& \sqrt{2}\eta\Omega_{\downarrow\uparrow}\frac{\hat\sigma_{\phi_s,1}-\hat\sigma_{\phi_s,2}}{2} \,\hat ae^{i\delta t+i\phi_m}+h.c., \label{eq:H-MS-RWA}
\end{align}
where we have defined the `spin' phase $\phi_s=(\phi_{2,\mathrm{r}}-\phi_{2,\mathrm{b}})/2$ and the `motion' phase $\phi_m^{(j)}=\phi_1+(\phi_{2,\mathrm{b}}+\phi_{2,\mathrm{r}})/2+\mathbf{\Delta k}\cdot\mathbf{r_j^{(eq)}}$, and the factor $1/\sqrt{2}$ comes from normalizing for the total power sent into the field $\mathbf{E_2}$. The operator $\hat\sigma_{\phi_s,j}=e^{i\phi_s}\hat\sigma_{-,j}+e^{-i\phi_s}\hat\sigma_{+,j}$. We assumed in the last line that the ion spacing has been chosen such that $\phi_m^{(j)}$ is the same on the two ions. As in the case of LS gates, this is ensured automatically by the placement of the ions at $y=0$ in RW1 but requires adjustment of the ion spacing along the trap axis in RW2.

In the SW configuration, we may write,
\begin{align}
\frac{\hat H_{\mathrm{int}}^{\mathrm{MS}}}{\hbar}=&\sum_j\Omega_{\downarrow\uparrow}\mathbf{k_1}\cdot\mathbf{r_j}\;\hat{\sigma}_{-,j}\nonumber \\
&\qquad\times\Big\{e^{i(\omega_{y_s}+\delta)t+i\phi^{(j)}}+e^{-i(\omega_{y_s}+\delta) t-i\phi'{}^{(j)}}\Big\} + h.c.
\end{align}

\begin{figure*}[btp]
\centering
\includegraphics[width=\linewidth]{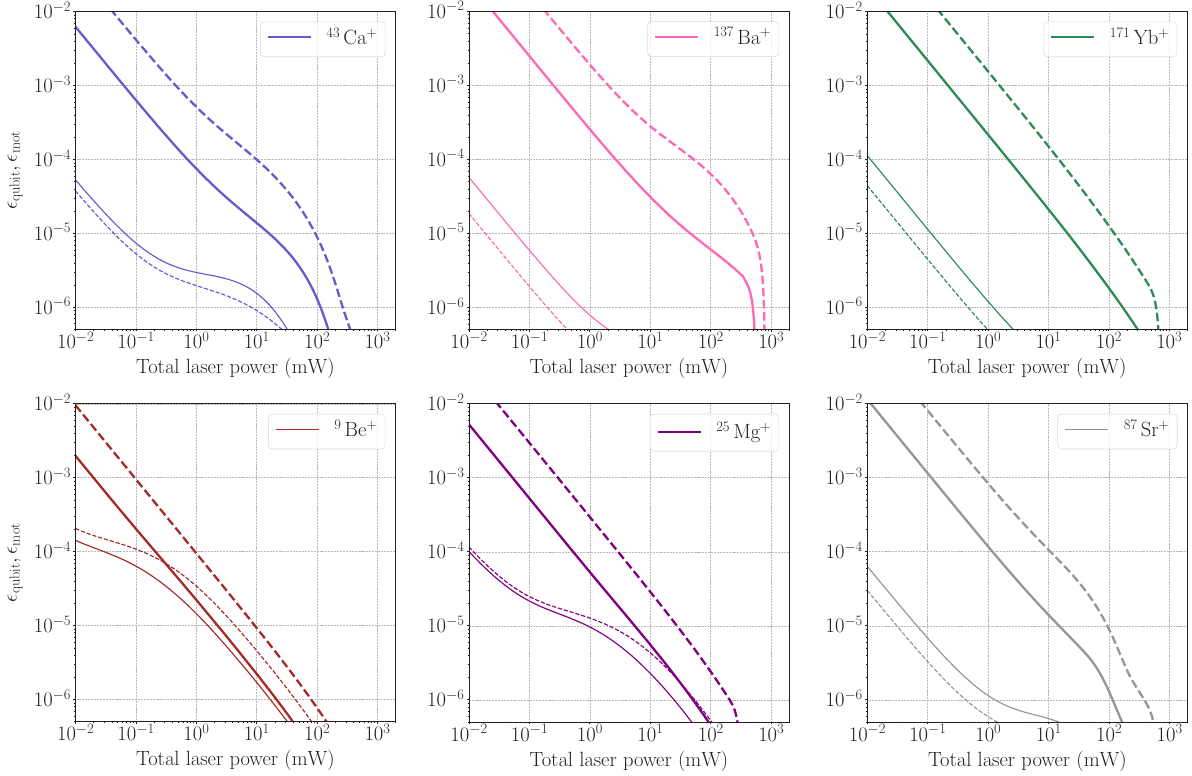}
\caption{Minimum achievable qubit decoherence and motional contributions to the MS gate error as a function of available total laser power. Solid (dashed) lines show calculated error in the SW (RW2) configuration. Thick (thin) lines denote the qubit decoherence (motional) contribution to the total gate error $\epsilon_{\mathrm{SPS}}$.}
\label{fig:MS_qubit_plus_recoil}
\end{figure*}
where we have neglected the correction to the coupling to the motion due to the beam propagating along $\mathbf{k_2}$. We define $\phi^{(j)}=-\phi_{2,\mathrm{b}}-\mathbf{k_2}\cdot\mathbf{r_j^{(eq)}}$ and $\phi'{}^{(j)}=-\phi_{2,\mathrm{r}}-\mathbf{k_2}\cdot\mathbf{r_j^{(eq)}}$. Retaining only the near-resonant terms in the above expression, we find that the gate Hamiltonian takes the form \eqref{eq:H-MS-RWA} with $\phi_s=(\phi_{2,\mathrm{r}}-\phi_{2,\mathrm{b}})/2$ as before and $\phi_m=(\phi_{2,\mathrm{r}}+\phi_{2,\mathrm{b}})/2-\mathbf{k_2}\cdot\mathbf{r_j^{(eq)}}$. We assume again that the ions are spaced along the trap axis by an integer number of wavelengths $\lambda_{\mathrm{eff}}=2\pi/\mathbf{k_2}\cdot\mathbf{u_x}$ in order to maximize the gate drive strength.

For $K$-loop gates of duration $\tau_{\mathrm{g}}$, the gate detuning is set to $\delta=2\pi K/\tau_{\mathrm{g}}$, and we have the following condition on the gate drive \cite{lee2005phase}:
\begin{equation}
\eta\Omega_{\mathrm{gate-drive}}\tau_{\mathrm{g}}=\pi\sqrt{K}, \label{eq:gate-drive-constraint}
\end{equation}
where $\Omega_{\mathrm{gate-drive}}=2|\Omega_{\downarrow\downarrow}|$ for the LS gate and $\Omega_{\mathrm{gate-drive}}=\sqrt{2}|\Omega_{\downarrow\uparrow}|$ for the MS gate. In this paper, we take $K=1$ throughout.

\subsection{SW-induced motional frequency shift}
We now consider the shift in the secular frequency introduced by the standing wave field in the SW configuration. We calculate the strength of the static AC Stark shift $\Omega_{\mathrm{SS},s}$ on qubit state $s$ from \eqref{eq:H_ss'}:
\begin{align}
\Omega_{\mathrm{SS},s} = g_1^2&\sum_e\Big(\frac{\bra{s}\hat{\mathbf{r}}_{\mathbf{el}}\cdot\hat{\epsilon}_i\ket{e}\bra{e}\hat{\mathbf{r}}_{\mathbf{el}}\cdot\hat{\epsilon}_i\ket{s}}{\mu^2(\omega_{eg}-\omega_{\mathrm{las}})}\nonumber\\
&\qquad\qquad+\frac{\bra{s}\hat{\mathbf{r}}_{\mathbf{el}}\cdot\hat{\epsilon}_i\ket{e}\bra{e}\hat{\mathbf{r}}_{\mathbf{el}}\cdot\hat{\epsilon}_i\ket{s}}{\mu^2(\omega_{eg}+\omega_{\mathrm{las}})}\Big).
\end{align}
For the large Raman detunings considered here these shifts on the two qubit levels are approximately equal for the Zeeman qubits, as well as for the clock qubits \cite{lee2005phase}, so we denote them simply by $\Omega_{\mathrm{SS}}$. The optical potential due to the SW then takes the form
\begin{align}
\frac{\hat H_{\mathrm{opt-pot}}}{\hbar} &= 2\Omega_{\mathrm{SS}}(\mathbf{k_1}\cdot\mathbf{u_y})^2(\hat{q}_{y,1}^2+\hat{q}_{y,2}^2) \nonumber\\
&= 2\Omega_{\mathrm{SS}}(\mathbf{k_1}\cdot\mathbf{u_y})^2\sum_{l=y_c,y_s}y_{l}^{(0)2}\left(\hat a_l e^{-i\omega_l t}+\hat a_l^\dagger e^{i\omega_l t}\right)^2 \nonumber \\
&= 2\Omega_{\mathrm{SS}}(\mathbf{k_1}\cdot\mathbf{u_y})^2\sum_{l=y_c,y_s}2y_{l}^{(0)2}\left(\hat a_l^\dagger \hat a_l+\frac{1}{2}\right)
\end{align}
Therefore, the frequency shift of the mode $y_s$ is,
\begin{equation}
\Delta\omega_{y_s}=4(\eta^{\mathrm{SW}})^2\,\Omega_{\mathrm{SS}}
\end{equation}
The frequency shift can be calculated exactly in the small-detuning limit $|\Delta|\ll\omega_{\mathrm{fine}}$, where it can be shown that $|\Omega_{\mathrm{SS}}|/|\Omega_{\downarrow\downarrow}|=E_1/E_2$. Then, using the constraint \eqref{eq:gate-drive-constraint} for the LS gate, and the optimal power distribution \eqref{eq:optimal-power-dist} in the low-detuning limit, we have, $\Delta\omega_{y_s}=\delta\sqrt{\eta^{\mathrm{SW}}/\bar\alpha}$.

The Stark shift term couples with the same phase to both polarization components of the Raman field, whereas the gate drive couples with equal magnitude and opposite sign to the left-circular and right-circular polarizations. Therefore, the frequency shift $\Delta\omega_{y_s}$ in the small-detnuning limit sets a lower bound. In Fig. \ref{fig:freq_shift}, we plot the frequency shift given a target gate error from spontaneous photon scattering $\epsilon_{\mathrm{SPS}}$ (see main text and Appendix \ref{App:scatter} for details on how this is calculated).

For relatively large gate errors (i.e. small Raman beam detunings $\Delta$ compared to $\omega_{\mathrm{fine}}$), the frequency shift remains close to its value in the small-detuning limit, and we assume that the difference frequency $\omega_2-\omega_1$ of the Raman beams is adjusted such that the gate detuning $\delta$ remains fixed. When $\Delta\omega_{y_s}/\omega_{y_s}\not\ll1$, eg. for the LS gate on ${}^{40}\mathrm{Ca}^+$ with $\epsilon_{\mathrm{SPS}}\lesssim10^{-5}$, corrections to the Lamb-Dicke parameter due to the frequency shift have to be taken into account. We have considered the `bare' mode frequencies $\omega_{y_c}$, $\omega_{y_s}$ in our calculations of the gate dynamics and the scattering error here, so these corrections will modify slightly the results presented in Figs. \ref{fig:LS_result} and \ref{fig:MS_result}. In such cases, adiabatic pulse-shaping on the gate pulses can be used to avoid squeezing the motional state when the gate beams are pulsed on \cite{robalo2025fast}. 

\section{Spontaneous photon scattering due to the SW} \label{App:scatter}

The rate of SPS due to RW fields and the associated error in two-qubit gates have been discussed in detail eg. in Refs. \cite{ozeri2007errors, sawyer2021wavelength, moore2023photon}. Scattering due to a SW has to be treated separately since the absorption from the laser field couples strongly to the motion.

Consider two ions illuminated by the SW field $\mathbf{E_1}(\mathbf{r},t)=\boldsymbol{\epsilon_1}\sqrt{2}E_1\sin(k_{||}y)\cos(\omega_{\mathrm{las}} t)$ as in the right panel of Fig.~\ref{fig:beam_geometries_schematic}. The scattering rate calculated from second-order perturbation theory, shown in eq.~\eqref{eq:scattering_rate_if} in the main text, was derived assuming that the coupling of the laser field to the external degrees of freedom is negligible. Here instead we calculate scattering rates $\Gamma_{(i,n_{y_c},n_{y_s})(f,n'_{y_c},n'_{y_s})}$, explicitly accounting for transitions between motional states. The SW field in our case has a gradient along the $y$-direction with zero field in the $y=0$ plane, so we only need to consider the COM and `stretch' modes in the $y$-direction. In the interaction picture with respect to the bare Hamiltonian of the ion, we write \cite{loudon2000quantum},
\begin{equation}
\Gamma^{(j)}_{(i,n_{y_c},n_{y_s})(f,n'_{y_c},n'_{y_s})} = 2g_1^2\sum_{\epsilon_{\mathrm{sc}}}\left | \sum_{e} \chi^{(e,\epsilon_{\mathrm{sc}}, j)}_{(i_j,n_{y_c},n_{y_s})(f_j,n'_{y_c},n'_{y_s})}\right |^2, \label{eq:scattering_rate_if_nn'}
\end{equation}
with the modified scattering amplitudes
\begin{align}
\chi^{(e,\epsilon_{\mathrm{sc}},j)}_{(i_j,n_{y_c},n_{y_s})(f_j,n'_{y_c},n'_{y_s})} = \chi^{(e,\epsilon_{\mathrm{sc}})}_{i_jf_j}\bra{n'_{y_c}n'_{y_s}}\sin(k_{||}\hat y_j)\ket{n^{}_{y_c}n^{}_{y_s}}, \label{eq:scattering-amp-if-nn'}
\end{align} 
where the scattering amplitude $\chi^{(e,\epsilon_{\mathrm{sc}})}_{i_jf_j}$ is defined in eq.~\eqref{eq:chi}. All Raman scattering events, where $\ket{f}\neq\ket{i}$, give rise to a gate error regardless of the motional state at the end of the scattering event \cite{ozeri2007errors}, so the Raman decoherence rate is proportional to,

\begin{widetext}
\begin{align}
\sum_j\sum_{f_j\neq i_j}\sum_{n'_{y_c},n'_{y_s}} \Gamma^{(j)}_{(i_j,n_{y_c},n_{y_s})(f_j,n'_{y_c},n'_{y_s})} \quad&=2g_1^2\sum_j\bra{n_{y_c},n_{y_s}}\sin^2(k_{||}\hat {q}_{y,j})\ket{n^{}_{y_c},n^{}_{y_s}}\sum_{\epsilon_{\mathrm{sc}},f_j\neq i_j}\left|\sum_e\chi^{(e,\epsilon_{\mathrm{sc}})}_{i_jf_j}\right|^2 \nonumber \\
&\approx g_1^2 k_{||}^2\sum_j\bra{n_{y_c},n_{y_s}} (\hat y_c + (-1)^j \hat y_s)^2\ket{n^{}_{y_c},n^{}_{y_s}}\sum_{\epsilon_{\mathrm{sc}},f_j\neq i_j}\left|\sum_e\chi^{(e,\epsilon_{\mathrm{sc}})}_{i_jf_j}\right|^2 \nonumber \\
&=g_1^2k_{||}^2\,\sum_l\bra{n_{l}}\hat y_l^2\ket{n_{l}}\sum_j\sum_{\epsilon_{\mathrm{sc}},f_j\neq i_j}\left|\sum_e\chi^{(e,\epsilon_{\mathrm{sc}})}_{i_jf_j}\right|^2 \nonumber \\
&\qquad\qquad+2g_1^2k_{||}^2\,\sum_j (-1)^j\bra{n_{y_c}}\hat y_c\ket{n_{y_c}}\bra{n_{y_s}}\hat y_s\ket{n_{y_s}}\sum_{\epsilon_{\mathrm{sc}},f_j\neq i_j}\left|\sum_e\chi^{(e,\epsilon_{\mathrm{sc}})}_{i_jf_j}\right|^2 \nonumber \\
&=g_1^2\sum_l \eta_l^2\left(2\bra{n_l}\hat{n}_l\ket{n_l}+1\right) \sum_j\sum_{\epsilon_{\mathrm{sc}},f_j\neq i_j}\left|\sum_e\chi^{(e,\epsilon_{\mathrm{sc}})}_{i_jf_j}\right|^2. \label{eq:scattering_rate_if_summed}
\end{align}
Here, we have expanded the ion displacements around $y=0$ in the basis of the motional modes and in going to the last line, defined $\eta_l\equiv k_{||}y_{l}^{(0)}$ and dropped the fast-rotating terms.

Since the motional state evolves during the gate, the scattering rates are time-dependent. We expand the displaced states $\ket{\alpha_{s_1s_2}^{(l)}(t)}$ in the Fock basis as,
\begin{equation}
\ket{\alpha_{s_1s_2}^{(l)}(t)} \equiv \sum_{n_l=0}^\infty c_{s_1s_2,n_l}(t) \ket{n_l},
\end{equation}
with $\sum_{n_l}|c_{s_1s_2,n_l}(t)|^2=1$. First consider the LS gate, where the coherent displacements are conditioned on the two-qubit state in the $z$-basis, so $s_j=i_j\in\{\downarrow,\uparrow\}$. We calculate the instantaneous Raman scattering-induced decoherence rate from the SW as a sum of rates~\eqref{eq:scattering_rate_if_summed} weighted by the Fock state occupancies in the initial state: 
\begin{align}
\Gamma_{\mathrm{Raman},1} &= g_1^2\sum_{i_1,i_2} p_{i_1i_2}\times \nonumber\\
&\quad\sum_l\eta_l^2\left(2\sum_{n_l}|c_{i_1i_2,n_l}|^2\bra{n_l}\hat n_l\ket{n_l}+1\right)\times\nonumber\\
&\qquad\qquad\qquad\qquad\qquad\sum_j\sum_{\epsilon_{\mathrm{sc}},f_j\neq i_j}\left|\sum_e\chi^{(e,\epsilon_{\mathrm{sc}})}_{i_jf_j}\right|^2 \nonumber \\
&= g_1^2\sum_{i_1,i_2} p_{i_1i_2}\sum_l\eta_l^2\left(2\bra{\alpha_{i_1i_2}^{(l)}}\hat n_l\ket{\alpha_{i_1i_2}^{(l)}}+1\right)\times \nonumber\\
&\qquad\qquad\qquad\qquad\qquad\sum_j\sum_{\epsilon_{\mathrm{sc}},f_j\neq i_j}\left|\sum_e\chi^{(e,\epsilon_{\mathrm{sc}})}_{i_jf_j}\right|^2 \nonumber \\
&= g_1^2\sum_{i_1,i_2} p_{i_1i_2}\sum_l\eta_l^2\left(2|\alpha_{i_1i_2}^{(l)}(t)|^2+1\right)\times \nonumber\\
&\qquad\qquad\qquad\qquad\qquad\sum_j\sum_{\epsilon_{\mathrm{sc}},f_j\neq i_j}\left|\sum_e\chi^{(e,\epsilon_{\mathrm{sc}})}_{i_jf_j}\right|^2.
\label{eq:decoherence_rate_Raman_SW}
\end{align}

We consider the decoherence rate averaged over all two-qubit initial states by taking $p_{i_1i_2}=1/4$. We simplify our expression further by expanding the sums over $j$ and $i_1i_2$:
\begin{align}
\Gamma_{\mathrm{Raman},1} = &g_1^2\sum_l\frac{1}{4}\eta_l^2\left(2|\alpha_{\downarrow\downarrow}^{(l)}(t)|^2+1\right)\times\sum_{\epsilon_{\mathrm{sc}}}\left(\sum_{f\neq \downarrow}\left|\sum_e\chi^{(e,\epsilon_{\mathrm{sc}})}_{\downarrow f}\right|^2 + \sum_{f\neq \downarrow}\left|\sum_e\chi^{(e,\epsilon_{\mathrm{sc}})}_{\downarrow f}\right|^2\right) \nonumber \\
&+g_1^2\sum_l\frac{1}{4}\eta_l^2\left(2|\alpha_{\downarrow\uparrow}^{(l)}(t)|^2+1\right)\sum_{\epsilon_{\mathrm{sc}}}\left(\sum_{f\neq \downarrow}\left|\sum_e\chi^{(e,\epsilon_{\mathrm{sc}})}_{\downarrow f}\right|^2 + \sum_{f\neq \uparrow}\left|\sum_e\chi^{(e,\epsilon_{\mathrm{sc}})}_{\uparrow f}\right|^2\right) \nonumber \\
&+g_1^2\sum_l\frac{1}{4}\eta_l^2\left(2|\alpha_{\uparrow\downarrow}^{(l)}(t)|^2+1\right)\sum_{\epsilon_{\mathrm{sc}}}\left(\sum_{f\neq \uparrow}\left|\sum_e\chi^{(e,\epsilon_{\mathrm{sc}})}_{\uparrow f}\right|^2 + \sum_{f\neq \downarrow}\left|\sum_e\chi^{(e,\epsilon_{\mathrm{sc}})}_{\downarrow f}\right|^2\right) \nonumber \\
&+g_1^2\sum_l\frac{1}{4}\eta_l^2\left(2|\alpha_{\uparrow\uparrow}^{(l)}(t)|^2+1\right)\sum_{\epsilon_{\mathrm{sc}}}\left(\sum_{f\neq \uparrow}\left|\sum_e\chi^{(e,\epsilon_{\mathrm{sc}})}_{\uparrow f}\right|^2 + \sum_{f\neq \uparrow}\left|\sum_e\chi^{(e,\epsilon_{\mathrm{sc}})}_{\uparrow f}\right|^2\right) \\
=&g_1^2\left(\sum_{i_1i_2}\sum_l\frac{1}{4}\eta_l^2\left(2|\alpha_{i_1i_2}^{(l)}(t)|^2+1\right)\right)\sum_{\epsilon_{\mathrm{sc}}}\sum_{i,f\neq i}\left|\sum_e\chi^{(e,\epsilon_{\mathrm{sc}})}_{i f}\right|^2 \label{eq:Gamma_Ramana_SW_big_boy}
\end{align}
\end{widetext}

\begin{figure}[t]
\centering
\includegraphics[width=0.82\linewidth]{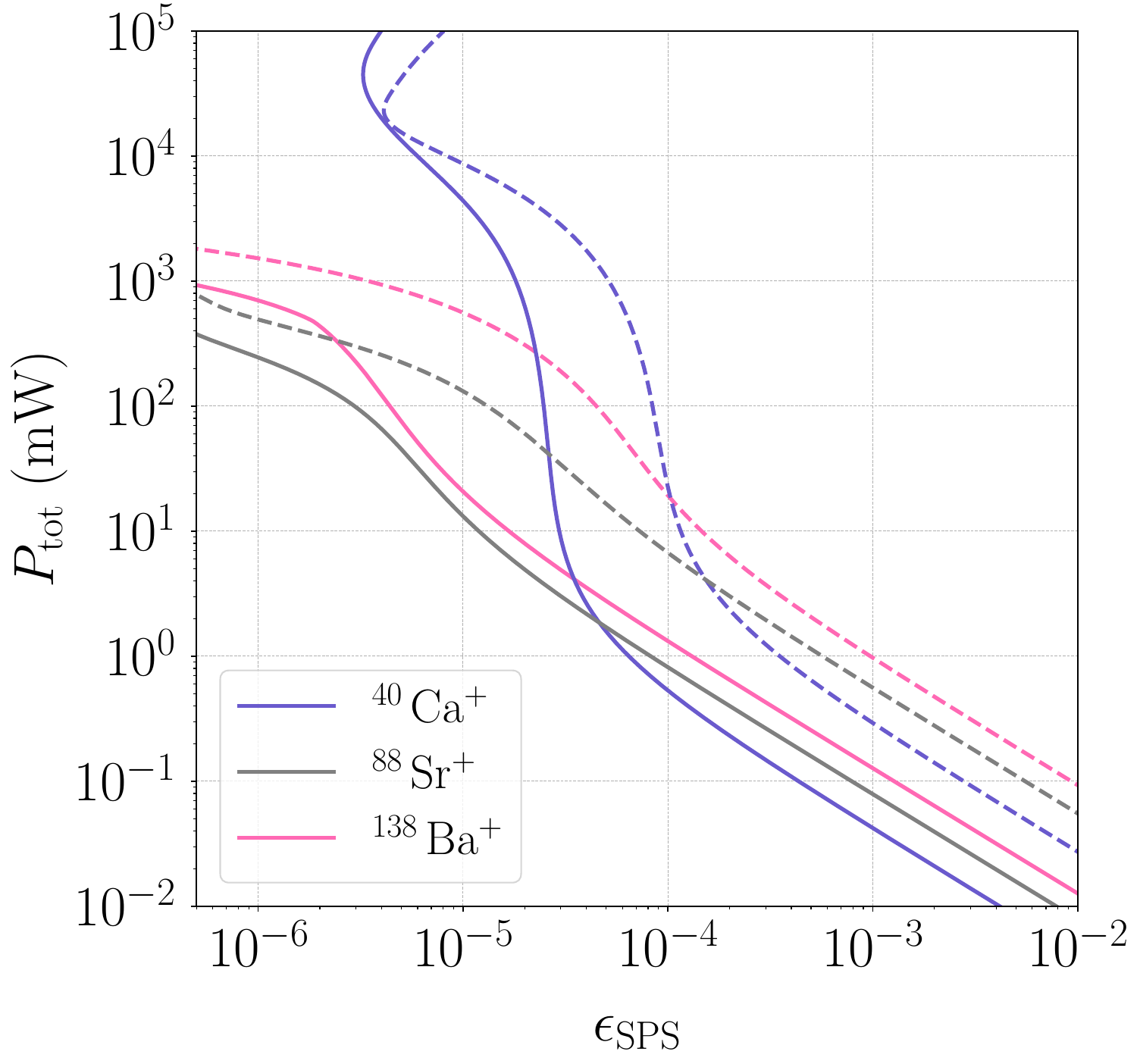}
\caption{Total laser power required given a target SPS-induced gate error $\epsilon_{\mathrm{SPS}}$, for the LS gate in the SW scheme (solid line) and the RW2 scheme (dashed line). The curve for the RW2 scheme is reproduced for comparison from Fig.~\ref{fig:LS_result}a, while that for the SW scheme is calculated assuming that every Rayleigh scattering event induces a gate error, showing that the predicted enhancement is minimally affected except at extremely low errors where Rayleigh scattering dominates.}
\label{fig:LS_qubit_plus_recoil_pessimistic}
\end{figure}
\begin{figure}[t]
\centering
\includegraphics[width=0.82\linewidth]{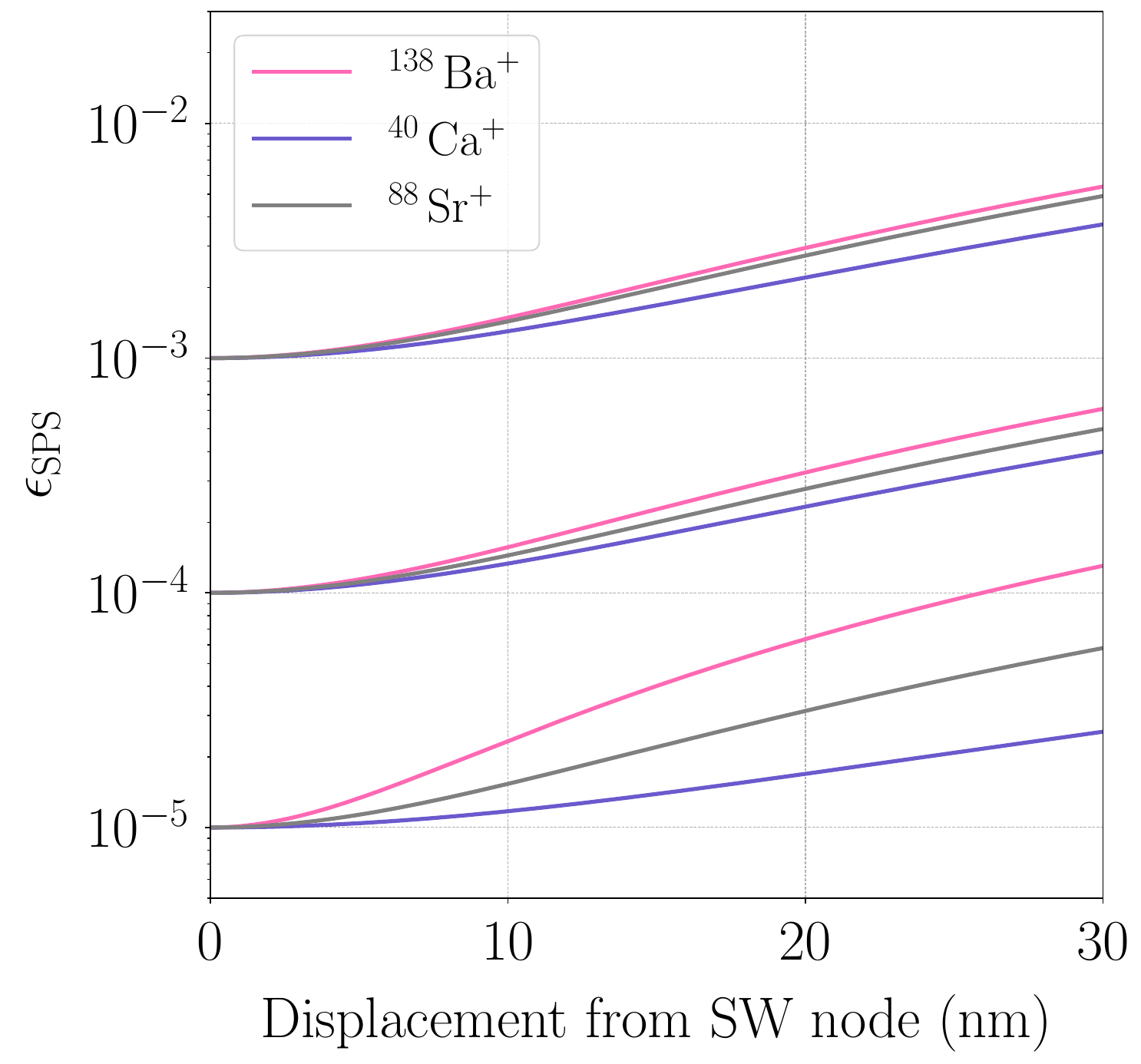}
\caption{SPS-induced gate error $\epsilon_{\mathrm{SPS}}$, for the LS gate in the SW scheme, for small displacements of the trap equilibrium position from the SW intensity node. Curves are presented for target SPS errors $10^{-3}$, $10^{-4}$ and $10^{-5}$. The excess error grows quadratically with the displacement, making the SW scheme robust to positioning imperfections anticipated in experiments.}
\label{fig:excess_error_due_to_positioning}
\end{figure}

For a one-loop gate, the coherent displacements, up to an overall phase, are \cite{langer2006high}:
\begin{align*}
\alpha_{\downarrow\downarrow}^{(y_s)}&=\alpha_{\uparrow\uparrow}^{(y_s)}=0, \\
\alpha_{\downarrow\uparrow}^{(y_s)}&=-\alpha_{\uparrow\downarrow}^{(y_s)}=\frac{e^{-i\delta t}-1}{2}, \\
\alpha_{i_1i_2}^{(y_c)}&=0,
\end{align*}
such that the time-averaged squared excursion on the driven states is,
\begin{equation}
\frac{1}{\tau_{\mathrm{g}}}\int_0^{\tau_{\mathrm{g}}}dt\,|\alpha_{\downarrow\uparrow}^{(y_s)}(t)|^2=\frac{1}{\tau_{\mathrm{g}}}\int_0^{\tau_{\mathrm{g}}}dt\,|\alpha_{\uparrow\downarrow}^{(y_s)}(t)|^2=\frac{1}{2}.
\end{equation}
The absorption rates in the other basis states and due to the spectator mode are constant in time. Then, in the SW configuration of Fig.~\ref{fig:beam_geometries_schematic}, we include the contribution of the RW beam and write the gate error due to Raman scattering as,
\begin{equation}
\epsilon_{\mathrm{SPS, Raman}} = \tau_{\mathrm{g}}\left(\eta^2\bar\alpha^2 g_1^2+ g_2^2\right)\sum_{i,f\neq i}\sum_{\epsilon_{\mathrm{sc}}}\left|\sum_e\chi^{(e,\epsilon_{\mathrm{sc}})}_{if}\right|^2. \label{eq:err-Raman-SW-App}
\end{equation}
where $\eta=\eta^{\mathrm{SW}}$ as defined in eq.~\eqref{eq:eta_SW} and in Appendix~\ref{App:gateham}. We have defined
\begin{align}
\bar\alpha^2 &\equiv \frac{1}{4\eta^2}\sum_{i_1i_2}\eta_l^2\left(2|\alpha_{i_1i_2}^{(l)}(t)|^2+1\right)  \\
&=\frac{\omega_{y_s}}{\omega_{y_c}}+\frac{3}{2}.
\end{align}
In MS gates, the coherent displacements are conditioned on the state in the $xy$-basis, with $s_1,s_2\in\{+_{\phi_{\mathrm{s}}},-_{\phi_{\mathrm{s}}}\}$. For equal average occupancies of the qubit levels, we can replace $|\sum_e\chi^{(e,\epsilon_{\mathrm{sc}})}_{i f}|^2$ by the average, $\frac{1}{2}\left|\sum_e\chi^{(e,\epsilon_{\mathrm{sc}})}_{\downarrow f}\right|^2+\frac{1}{2}\left|\sum_e\chi^{(e,\epsilon_{\mathrm{sc}})}_{\uparrow f}\right|^2$.
Then, in analogy to eq.~\eqref{eq:Gamma_Ramana_SW_big_boy}, we may write,
\begin{align}
\Gamma_{\mathrm{Raman},1}=&g_1^2\left(\sum_{s_1s_2}\sum_l\frac{1}{4}\eta_l^2\left(2|\alpha_{s_1s_2}^{(l)}(t)|^2+1\right)\right)\times\nonumber\\
&\qquad\qquad\qquad\qquad\quad\sum_{\epsilon_{\mathrm{sc}}}\sum_{i,f\neq i}\left|\sum_e\chi^{(e,\epsilon_{\mathrm{sc}})}_{i f}\right|^2,
\end{align}
and therefore, the expression~\eqref{eq:err-Raman-SW-App} applies directly to the MS gate.  Supplementing Fig.~\ref{fig:MS_result} in the main text, Fig.~\ref{fig:App_MS_result} presents our results for the MS gate for ${}^{9}\mathrm{Be}^+$, ${}^{25}\mathrm{Mg}^+$ and ${}^{87}\mathrm{Sr}^+$. We also plot the qubit decoherence and motional error contributions to SPS separately for each species, for the LS gate in Fig.~\ref{fig:LS_qubit_plus_recoil} and for the MS gate in Fig.~\ref{fig:MS_qubit_plus_recoil}.

In the main text, we calculate the Rayleigh scattering error using the average excitation rate $\left(\eta^2\bar\alpha^2 g_1^2+ g_2^2\right)$, as in the case of Raman scattering. A more careful treatment of the Rayleigh scattering-induced decoherence needs to fully account for spin-motion entanglement during the gate, which may modify the excitation rate via interference effects between the motional states. To show that the advantages predicted in the main text persist in a model with a more pessimistic error estimate from Rayleigh scattering, we consider a model where we calculate $\epsilon_{\mathrm{SPS, Rayleigh}}$ in the SW scheme by assuming that all Rayleigh scattering events due to the SW field result in a gate error. The Rayleigh decoherence rate due to RW fields is set by the squared difference of scattering amplitudes, as in the main text.

In Fig.~\ref{fig:LS_qubit_plus_recoil_pessimistic}, we plot the total laser power required to achieve a target gate error $\epsilon_{\mathrm{SPS}}$ for the LS gate. Comparing this to Fig.~\ref{fig:LS_result}a, we see that the qualitative features of the curve for the SW configuration as well as the predicted power advantage remain largely unaltered except at extremely low target gate errors, where Rayleigh scattering-induced errors are the dominant contribution.

Our analysis of the SW scheme thus far assumes that the trapping potential minium is precisely at a SW node. Small errors in ion positioning minimally alter gate dynamics due to the linearity of the field profile around the SW nodes, but lead to excess photon scattering from the SW field. The excess error is proportional to the SW intensity at the equilibrium trap position and therefore grows quadratically with the trap displacement from the node. In Fig. \ref{fig:excess_error_due_to_positioning}, we plot the estimated SPS error for the LS gate as a function of single-ion $y$-displacement $y_j^{\mathrm{(eq)}}$ in \eqref{eq:ion_position_operator}. We see that gate fidelities are robust to positioning inaccuracies of order $10$ nm, as have been measured in recent experiments \cite{vasquez2023control, clements2026sub, corsetti2026integrated, xing2025rapid} utilizing integrated addressing.  

Code for producing the figures in this paper can be found in Ref. \cite{repo}.

\bibliography{2qgatebib}

\end{document}